\documentclass[twocolumn]{aastex701}
\usepackage{amsmath,amstext,hyperref}
\usepackage[T1]{fontenc}
\usepackage{txfonts} 
\usepackage[figure,figure*]{hypcap} 

\newcommand*{\kms}{\,km~s$^{-1}$}
\newcommand{\Msun}{\mbox{\,$M_{\odot}$}}

\newcommand{\nmin}{10~} 
\newcommand{\nvobj}{67~} 

\shorttitle{DEIMOS}
\shortauthors{Geha}

\begin{document}

\title{The Keck/DEIMOS Stellar Archive: II. \\ Dynamical Masses and Metallicities for a Uniform Sample of Milky Way Satellites}

\author[0000-0002-7007-9725]{Marla~Geha}
\affiliation{Department of Astronomy, Yale University, New Haven, CT 06520, USA}
\email[show]{marla.geha@yale.edu}

\begin{abstract}

Population-level studies of Milky Way satellites used to constrain dark matter or the threshold of galaxy formation often rely on velocity dispersions and metallicities derived from heterogeneous spectroscopic analyses. Systematic differences between data reduction pipelines and membership criteria can masquerade as astrophysical signals, or obscure real trends.  Here, we present the largest self-consistent sample of spectroscopically-derived quantities for Milky Way satellite galaxies and globular clusters based on a homogeneous re-analysis of individual stars observed with the Keck/DEIMOS spectrograph.   We determine enclosed dynamical masses, mean [Fe/H] metallicities, and metallicity dispersions for 67 systems with 10 or more member stars.  At a given stellar mass, systems classified as satellite galaxies are well separated from globular clusters in their dynamical mass and mass-to-light ratios.   The average enclosed mass densities of satellite galaxies agree with semi-analytic CDM model predictions.  For satellite galaxies, we observe a break in the stellar mass-metallicity relation near $\log M_\star/M_\odot =4$ ($M_V\sim -4.5)$. Above this stellar mass, satellite galaxies show the well-known tight trend (0.16\,dex scatter in [Fe/H]) of decreasing metallicity with stellar mass; below $\log M_\star/ M_\odot=4$, the mass-metallicity relation flattens and/or increases in scatter. Satellite galaxies have internal metallicity scatter between 0.3-0.4\,dex across our stellar mass range.    These uniform measurements will enable tighter constraints on the stellar mass-halo mass relation, improved J-factor estimates for dark matter searches, and lay a foundation for interpreting the flood of new Milky Way satellites expected in the LSST/Roman/Euclid era.

\end{abstract}

\section{Introduction}

As a system, the Milky Way's stellar satellites provide a key snapshot of both galaxy and star cluster formation over a wide range of stellar mass.   For satellite galaxies, strong correlations exist between a galaxy's total stellar mass, dynamical/halo mass, and mean metallicity \citep[e.g.,][]{strigari2008, kirby2013,bullock2017,simon2019, sales2022, review2022}, which differ in comparison to globular or star clusters at the same luminosity \citep{Carretta2009,Baumgardt2018,pace2024}.   Both the relationship between quantities, and the scatter of systems around these scaling laws, provide insight into the formation processes of stellar systems, particularly at low stellar masses  \citep[e.g.,][]{wetzel2016,agetz2020,munshi2021, nadler2024A,ahvazi2025, go2025,taylor2025, brown2025,rey2025}.

The utility of the Milky Way's satellites depends, in part, on accurate and uniform estimates of individual stars’ kinematics and chemical abundances.  These properties have been determined for millions of bright stars via measurements from the {\it Gaia} satellite \citep{GaiaDR3}, and ground-based wide-field spectroscopic surveys including SDSS-SEGUE \citep{segue2008}, APOGEE \citep{Majewski2017}, LAMOST \citep{lamost2019}, and the DESI Survey \citep{koposov2025}.   Yet the limiting magnitude of current wide-field spectroscopic surveys ($r < 20$) means few stars are beyond the Milky Way Galaxy itself.   

Measuring kinematics and chemistries of individual stars in the Milky Way's satellites requires large aperture multi-object spectrographs and medium resolution spectroscopy, due to both larger distances and the higher spectral resolution needed to resolve lower dynamical masses as compared to the Milky Way.  Nonetheless, the number of radial velocity measurements for stars in Milky Way satellites is rapidly growing.  Many teams have contributed to the measurements of radial velocity data in Milky Way satellite galaxies \citep[e.g.,][]{Helmi2006,simon07a,Battaglia2008,li2019, pace2020,buttry2022,Tolstoy2023,walker2023,yang2025} and globular clusters \citep[e.g.,][]{Baumgardt2018,pace2023,Leitinger2025}.   The Deep Extragalactic Imaging Multi-Object Spectrograph \citep[DEIMOS; ][]{faber03a} on the Keck-II 10-meter telescope has contributed significantly to this effort, however, the data have been reduced and analyzed using a variety of different methods,
leading to a highly heterogeneous dataset \citep[e.g.,][]{simon07a,Martin2007, geha2009, kirby10a,willman2011,kim2016, collins2017, Longeard2020,2023A&A...672A.131A, cerny2024, Tan2025}.  The Keck Observatory Archive has recently opened access to over 20 years of archival DEIMOS data including hundreds of pointings towards Milky Way satellite galaxies and globular clusters.    

In \citet[Paper\,I;][]{geha_paper1}, we describe work in which we homogeneously reduce DEIMOS multi-slit spectra and determine stellar radial velocities and equivalent width-based metallicities for over 22,000 stars in 78 Milky Way satellite systems.  In this paper, we use this uniform dataset to determine integrated properties of Milky Way satellites observed with DEIMOS, including internal velocity dispersions, enclosed dynamical masses, mean metallicities and internal metallicity spreads.   We then assess correlations between properties for both satellite galaxies and globular/star clusters.  In \S\,\ref{sec_data}, we describe the DEIMOS sample and dataset.  In \S\,\ref{sec_mass}, we focus on kinematics, determining enclosed dynamical masses, mass-to-light ratios, and exploring the extent to which tidal forces from the Milky Way may be affecting our measurements.  In \S\,\ref{sec_metals}, we determine mean metallicity and internal metallicity spreads.   In \S\,\ref{sec:discussion}, we combine our kinematic and metallicity measurements, with the goal of better separating satellite
galaxies from star clusters, and then focus on satellite galaxies as probes of dark matter.  A comparison to literature measurements and selected predictions is provided in the Appendix.

\begin{figure*}[th!]
 \includegraphics[width=1.0\textwidth]{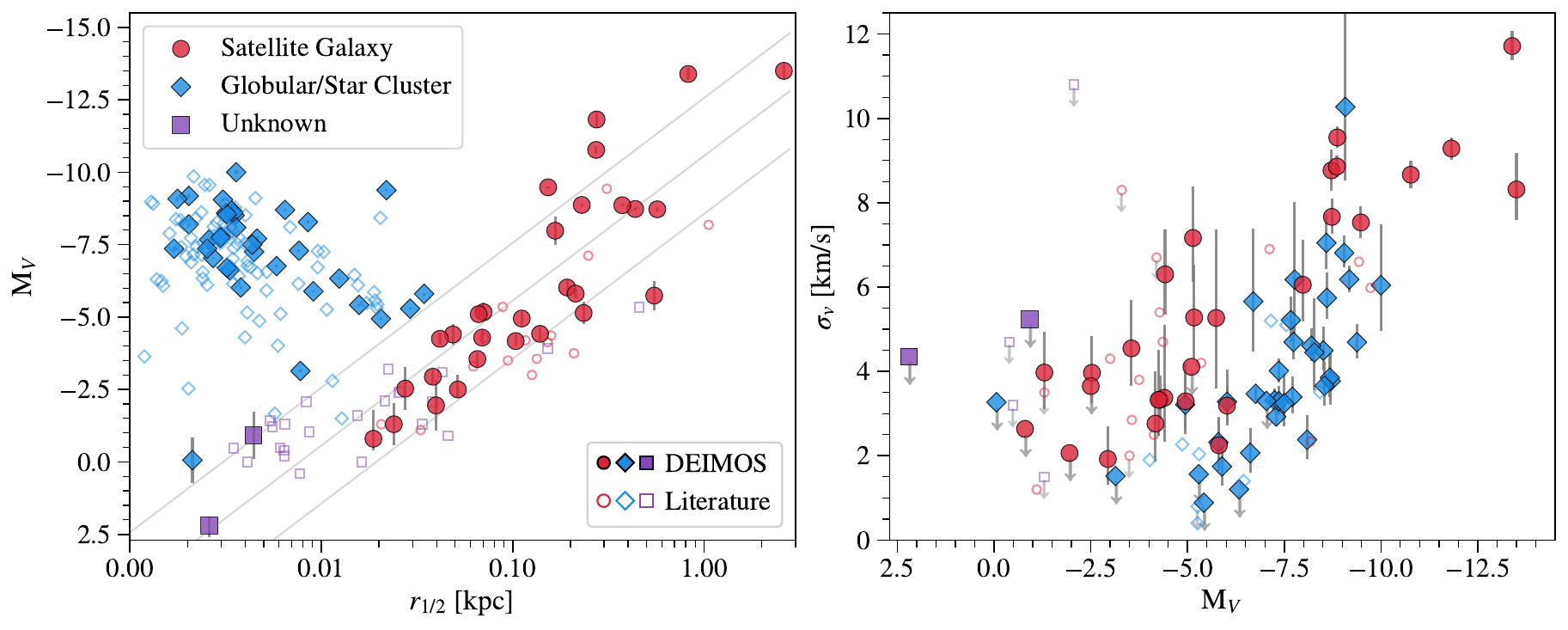}
\caption{{\it Left:\/} Absolute magnitude ($M_V$) versus half-light radius ($r_{\rm 1/2}$) for confirmed Milky Way satellite galaxies (red circles), globular/star clusters (blue diamonds) and ambiguously classified objects (purple squares), as compiled in v1.0.6 of \citet{pace2024}.   Solid symbols are systems with uniform DEIMOS measurements included in this work; open symbols indicate Milky Way systems without DEIMOS coverage.  Thin diagonal lines of constant effective surface brightness are shown from top to bottom: 26, 28, 30\, mag arcsec$^{-2}$.  {\it Right:\/}  Velocity dispersion ($\sigma_v$) versus $M_V$ as measured by this work (solid symbols), and system with non-DEIMOS literature measurements (open symbols).    While the DEIMOS sample represents a fraction of all known MW systems ({\it left}), it includes the majority of systems with measured internal velocity dispersions ({\it right}).  \label{fig_basic}}
\end{figure*}

\section{A Homogeneous Dataset}\label{sec_data}

Paper I \citep{geha_paper1} presents a uniform analysis of individual stars in Milky Way satellites as observed with the Keck II telescope and DEIMOS spectrograph \citep{faber03a}.  This currently represents the largest sample of Milky Way satellites with homogeneously reduced kinematics and metallicities of individual stars.   We include only data taken with the 1200G grating,  the highest available spectral resolution with DEIMOS, which provides a resolving power of R $\sim 6000$ across a wavelength range of $6500 - 8900\mbox{\AA}$.   As described in Paper I, multi-slit DEIMOS data spanning two decades were downloaded from the Keck Observatory Archive\footnote{https://koa.ipac.caltech.edu}\,(KOA).   Raw data are reduced to one-dimensional spectra using {\tt PypeIt} which provides near Poisson statistics-level sky subtraction \citep{pypeit2020}.  Radial velocities are determined via {\tt dmost},  a forward modeling method presented in Paper I, combining both synthetic telluric and stellar templates to determine stellar radial velocities.  The precision and accuracy of these measurements and measurement errors are assessed based on thousands of independent repeat observations and comparison to larger all-sky spectroscopic surveys. The velocity error floor for the {\tt dmost} pipeline is 1.1\kms.  The {\tt dmost} pipeline also determines stellar metallicities ([Fe/H]) based on the Ca II triplet, with a metallicity error floor of 0.1\,dex per star.  

We use only data from Paper I to maintain homogeneity, avoiding possible offsets or biases when combining datasets from different telescopes \citep[e.g.,][]{Tsantaki2022}.  In Figure~1, we plot the Milky Way satellites included in this work as solid symbols; open symbols are systems without archival DEIMOS coverage as compiled by \citet{pace2024}.  The majority of the solid DEIMOS points also have literature values, which we compare in Appendix A.   Our velocity dispersions generally agree with literature values, but have 20-50\% smaller errors (see Paper I, Figure 18).  Membership probabilities ($P_{\rm mem}$) are determined for individual stars in Paper I based on criteria in color-magnitude, velocity, equivalent widths, and where possible, {\it Gaia} DR3 parallax and proper motion measurements.  We adopt a membership threshold of $P_{\rm mem} > 0.5$.    For velocity-based quantities, we additionally remove stars identified as velocity variables (the $P_{\rm mem\_novar}$ sample from Paper I).  For [Fe/H]-based quantities, we remove stars fainter than $M_V > 3$ which lie outside of the empirical equivalent width-based metallicity calibration, as described in \S\,\ref{sec_metals}.  

The number of member stars per system with DEIMOS measurements varies significantly, from just 3 to over 1000 member stars (Table~\ref{table_objects}, see also Figure~16 in Paper\,I ).  There is also significant variation in radial coverage.  For the majority of systems, stellar members are measured between 0.1 to 2.5 effective half-light radii ($r_{\rm 1/2}$).   However, this region is entirely excluded for a handful of globular clusters due to crowding.  To address variation between systems, we explicitly state the median radius of the stellar sample used for each measurement.   Furthermore, we
report quantities only when \nmin or more member stars are measured in a given system.

 Throughout this paper, we plot systems as \textquotedblleft Satellite Galaxies" (red circles), \textquotedblleft Globular/Star Clusters" (blue diamonds), and systems where this identification is ambiguous or \textquotedblleft Unknown" (purple squares).  We use the published classifications compiled in v1.0.6 of \citet{pace2024}, which are based on a mix of photometric, kinematic, and metallicity information.   Our analysis below confirms these classifications.   Photometric quantities, including half-light radius,  total luminosity, stellar mass, and distances are also taken from \citet{pace2024}.   We note that stellar mass ($M_\star$) is computed directly from $M_V$ assuming a mass-to-light ratio of 2, appropriate for an old, metal-poor stellar population \citep{fsps2010}.

In this work, we explore Milky Way satellites as a population, examining trends between measured properties.  Shown in Figure~\ref{fig_basic}, our DEIMOS sample is representative of known Milky Way satellites.  The DEIMOS sample is complete for confirmed satellite galaxies with Declination $>-40^{\circ}$, with the exception of Antlia\,II and Crater\,II.  There are 12 additional confirmed satellite galaxies in the Southern hemisphere (${\rm Declination} < -40^{\circ}$), and thus not visible from the Keck telescope latitude. The DEIMOS sample include less than half of known Milky Way globular clusters.   A significant number of \textquotedblleft Unknown" systems shown in Figure~\ref{fig_basic} have recent DEIMOS data which will be published in Cerny et al.~(2026) using the same data reduction pipeline described in Paper\,I.   Finally, we note that the current Milky Way satellite sample itself is likely biased against extremely low surface brightness systems ($\mu > 31$\,mag arcsec$^{-2}$, left panel Figure~\ref{fig_basic}).  Upcoming deep imaging surveys will reveal whether or not such extreme systems exist in abundance, and to what extent this affects trends seen in the present Milky Way sample \citep[e.g.,][]{wheeler2025}.

\section{Dynamical Mass and Mass-to-Light Ratios}\label{sec_mass}

We focus first on the internal kinematics of Milky Way satellites, determining velocity dispersions based on individual stars (\S\,\ref{ssec_vdisp}), and then calculating the enclosed dynamical mass and mass-to-light ratios (\S\,\ref{ssec_mass_ml}).   We assume systems are in dynamical equilibrium, and test the validity of this assumption by examining the enclosed mass density (\S\,\ref{ssec:tides}).
Imposing the membership and radial cuts described below, we present \nvobj Milky Way systems (29 satellites galaxies and 38 globular clusters/ambiguous systems) with DEIMOS-only kinematic measurements.

\subsection{Velocity Dispersions}\label{ssec_vdisp}

With the goal of a uniform comparison, we determine the internal velocity dispersion for satellite galaxies using stars enclosed within two half-light radii ($2\,r_{\rm 1/2}$).  This radius was chosen to maximize the number of member stars, while minimizing the risk of including unbound or foreground Milky Way stars.   Since the velocity dispersions of Milky Way satellite galaxies are largely flat with radius \citep[e.g.,][]{Walker2007,battaglia2013}, this choice does not significantly affect the resulting velocity dispersions.   On the other hand, the velocity dispersions of globular clusters are a strongly decreasing function of increasing radius \citep[e.g.,][]{Baumgardt2018,sollima2019}, and the radial coverage for these systems often excludes the inner half-light radius due to crowding.   Thus, for globular clusters and ambiguous systems, we also attempt to measure the velocity dispersion within $2\,r_{\rm 1/2}$, but extend radial coverage in cases when less than 20\% of members are enclosed.  For all systems, we provide the median radius of member stars used to infer the velocity dispersion (Table~\ref{table_objects}).

To minimize the presence of unresolved binary stars or other velocity-variable stars, we  leverage multi-epoch measurements across the full DEIMOS archive.   We identify stars with variable velocities as described in Paper\,I, Section 5.2.  Only a fraction of the full sample has multi-epoch observations (20\% of member stars in satellite galaxies and 10\% of globular cluster stars).  Thus, there are certainly binaries present in the current sample which may affect the inferred velocity dispersions \citep[e.g.,][]{minor2019, pianta2022, Gration2025}.  Paper\,I provides secure first epoch velocity measurements to help identify binaries and other velocity-variable stars in future spectroscopic surveys.

We calculate a system's mean radial velocity ($v_{\rm sys}$) and internal velocity dispersion ($\sigma_{\rm v}$) using a two-parameter Gaussian likelihood function \citep[Eq 7-8; ][]{Walker2006}.  We use the {\tt dynesty} nested-sampling algorithm \citep{DYNESTY2020} to sample the posterior distribution, as it provides an estimate of the Bayesian evidence for each model.    We adopt flat (linear) priors in velocity and velocity dispersion, but explore whether logarithmic priors over the same region in velocity dispersion space affect our results.  For systems with 50 or more stars, the choice of prior does not affect the inferred velocity dispersion.  For smaller samples, the choice of prior is increasingly important: for 25 stars, the adopted prior affects the resulting DEIMOS dispersions up to 0.25\kms, while samples between 10-25 stars can differ as much as 0.75\kms.  A more thorough exploration on the choice of priors will be presented a future paper.   Here, we adopt a linear prior which produces more conservative (larger) upper limits in the limit of small sample size.   Nonetheless, in Paper\,I, we showed that these DEIMOS-derived velocity dispersions generally agree with literature values to within 1$\sigma$, but are more precise with 20-50\% smaller errors (see Figure~18 of Paper\,I).

In order to determine whether a system's internal velocity dispersion is resolved by the DEIMOS velocity sample, we compare the Bayesian evidence between our two-parameter Gaussian likelihood to a Gaussian model with zero internal dispersion (e.g., the observed velocity dispersion is due to measurement errors only).  Following recommended thresholds from \citet{Kass95}, we consider the velocity dispersion as resolved (non-zero) if the log of the ratio of evidences is larger than unity.   If resolved, we quote the 1-sigma errors (16th–84th percentile range) of the posterior distributions.   In cases when the dispersion is not resolved, we report only an upper limit on the velocity dispersion as the 95th percentile of the posterior distribution.

In the right panel of Figure~\ref{fig_basic}, we show the measured velocity dispersion versus absolute magnitude ($\sigma_v$ vs.~$M_V$).   Systems in which the velocity dispersion is resolved are plotted with 1-sigma error bars, while systems with unresolved dispersions are plotted with downward errors at the location of the 95\% upper confidence limit.   12 of \nvobj\ systems have unresolved velocity dispersions in the DEIMOS sample (see Table~1).   The smallest resolved velocity dispersion is the globular cluster NGC\,7492 ($\sigma_v = 1.7^{+0.5}_{-0.4}$\kms, 21 stars), while the strongest upper limit is the globular cluster Eridanus ($\sigma_v<0.9$\kms, 24 stars).  For both satellites galaxies and star clusters in Figure~\ref{fig_basic}, velocity dispersions loosely decrease with luminosity.   In \S\,\ref{ssec:DM}, we further explore the velocity dispersions of satellite galaxies, focusing on trends with stellar mass and half-light radius as constraints on dark matter models.

\begin{figure*}[th!]
 \includegraphics[width=1.0\textwidth]{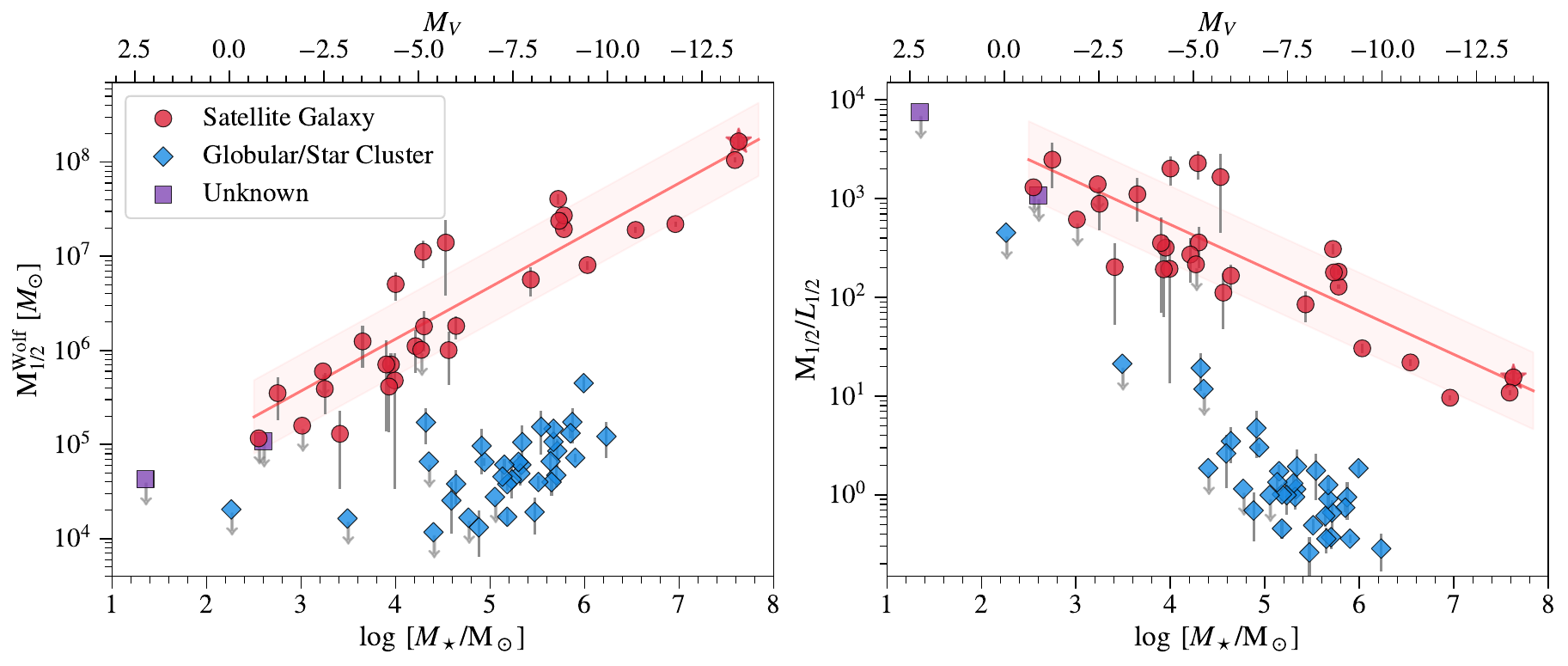}
\caption{{\it Left:\/} Dynamical mass enclosed within the half-light radius ($M_{\rm 1/2}$) versus stellar mass ($\log M_\star$).   {\it Right:\/}  Mass-to-light ratio (M$_{\rm 1/2} / L_{1/2}$) versus stellar mass.  In both panels, the top x-axis shows absolute magnitude ($M_V$).   At a given stellar mass, satellite galaxies (red circles) have larger dynamical mass and higher mass-to-light ratios compared to globular clusters (blue diamonds).  Globular clusters are consistent with baryonic-only mass.   Satellite galaxies show decreasing enclosed dynamical mass and increasing M$_{\rm 1/2} / L_{1/2}$ with decreasing stellar mass.   
\label{fig_wolf}}
\end{figure*}

\subsection{Enclosed Dynamical Mass and M/L}\label{ssec_mass_ml}

Given the wide range of physical properties in our sample, we opt for a simple estimate of  the enclosed dynamical mass.   We defer  more detailed mass modeling \citep[e.g.,][]{read2021,guerra2023,gravsphere2025} to future work.

\citet{wolf2010} showed that the dynamical mass enclosed within the half-light radius of a dispersion-supported system is robustly determined with only mild assumptions about the orbital anisotropy of its constituent stars.    We determine the dynamical mass according to \citet{wolf2010} as:  $M_{1/2} = 4\sigma_v^2r_{\rm 1/2}/G$, combining our DEIMOS velocity dispersions ($\sigma_v$) with the half-light  radius ($r_{\rm 1/2}$) from the literature compilation of \citet{pace2024}.   We compute errors on $M_{1/2}$ by Monte Carlo sampling errors in velocity dispersion and half-light radius. 

In the left panel of Figure~\ref{fig_wolf}, we plot the dynamical mass as a function of stellar mass, $M_{1/2}$ vs.~$M_{\star}$.   There is a general trend of decreasing  dynamical mass with decreasing stellar mass for both globular clusters and satellite galaxies.  However as expected, satellite galaxies have significantly larger dynamical mass at a given stellar mass.  The DEIMOS spectral resolution limits the minimum measured dynamical mass above $\log[M_{1/2}/M_\odot] \sim 4$.  

In the right panel of Figure~\ref{fig_wolf}, we plot the enclosed mass-to-light ratio (M$_{\rm 1/2} / L_{1/2}$) as a function of stellar mass, where $L_{1/2}$ is half the total $V-$band luminosity.   For globular clusters, we expect M$_{\rm 1/2} / L_{1/2}= 1.5 -3$, 
consistent with baryonic-only old stellar populations \citep{fsps2010}.   There is one globular cluster, Sgr\,II ($\log M_\star /M_\odot = 4.5$,  M$_{\rm 1/2} / L_{1/2} = 19\pm8$), which appears to lie between the globular cluster and satellite galaxy population.  Our velocity dispersion for Sgr\,II of $2.3\pm0.5$\,\kms\ (34 stars) is slightly larger than that of \citet{Longeard2021}, $1.7\pm0.5$\,\kms\ (47 stars), based on VLT FLAMES data.   Both samples suggest a very small metallicity dispersion, and we thus maintain that this object is correctly classified as a star cluster (see also \citealt{Zaremba2025}).   While a handful of globular clusters have large upper M$_{\rm 1/2} / L_{1/2}$ limits, these are all faint systems ($\log M_\star /M_\odot < 4.5$), whose expected velocity dispersions are below the DEIMOS velocity uncertainty floor.  Nonetheless, the median resolved mass-to-light ratio for all DEIMOS globular clusters is M$_{\rm 1/2} / L_{1/2} = 1.0$.  This is slightly lower than predicted for a baryon-only system because we are determining the average  velocity dispersions enclosed within 2$r_{\rm 1/2}$, rather than fitting a model to the velocity dispersion profile.

On the other hand, for satellite galaxies in Figure~\ref{fig_wolf}, the mass-to-light ratio clearly increases with decreasing stellar mass.  For luminous satellite galaxies ($\log M_\star /M_\odot > 6$), mass-to-light ratios are well in excess of the stellar mass alone: M$_{\rm 1/2} / L_{1/2} \sim 10$ -- $30$, with a minimum value for Leo\, I of M$_{\rm 1/2} / L_{1/2} = 10\pm1$.   For the least luminous satellite galaxies ($\log M_\star /M_\odot < 5$) mass-to-light ratios rise above 1000, with a maximum value for Segue\,1 of M$_{\rm 1/2} / L_{1/2}\sim 2500 \pm 1200$.  In both panels of Figure~\ref{fig_wolf}, we fit a linear function for satellite galaxies only.   We include only systems with resolved velocity dispersions, showing the best-fit model and rms scatter in the y-direction.  We first fit the dynamical masses versus stellar mass, finding $M_{1/2} = 0.54^{+0.07}_{-0.03} \log\,M_{\star} + 3.89^{+0.27}_{-0.29}$.   The slope of this relationship is in good agreement with predicted slope of 0.55 from the GRUMPY semi-analytical model of galaxy formation applied to the Caterpillar dark matter-only simulations \citep[Eq.~8;][]{Kravtsov2023}, which we overplot in Figure~\ref{fig_appendix_mass} and discuss further in \S\,\ref{ssec:DM}.

In the right panel of Figure~\ref{fig_wolf}, we also fit a linear model to M$_{\rm 1/2} / L_{1/2}$ vs.~$\log M_\star$, again only for satellite galaxies with resolved velocity dispersions.  Recent numerical simulations suggest a minimum stellar mass for galaxy formation, and that extremely low stellar mass systems are hosted by a wide range of dark matter halo masses (e.g., the stellar mass-halo mass relationship has a large scatter).  Suggested stellar mass thresholds for galaxy formation range from a thousand \citep{taylor2025,brown2025} to a few hundred \citep{munshi2021,wheeler2025b} solar masses.   However, these prediction remain highly dependent on differences in baryonic physics prescriptions such as gas heating/cooling physics, reionization, and molecular hydrogen cooling models \citep[e.g.,][]{nadler2025}.   We note a lack of extremely high mass-to-light ratio systems at the faintest end of this relationship ($\log M_\star /M_\odot < 3.5$, M$_{\rm 1/2} / L_{1/2}$ > 2500) which would be easily detected if present.   While the lack of high M/L systems might imply that the scatter in the stellar mass-halo mass relation does not significantly increase at extremely low stellar masses \citep{nadler2020,daneli2023}, there are also a number of systems with M$_{\rm 1/2} / L_{1/2}$ upper limits.   If the true M$_{\rm 1/2} / L_{1/2}$ of these unresolved systems are significantly smaller than current limits, this would increase scatter below $\log M_\star /M_\odot < 3.5$ as compared to more luminous Milky Way satellite galaxies.   Resolving this issue requires larger samples of individual stars with higher spectral resolution than possible with the Keck/DEIMOS spectrograph.

\begin{figure*}[t!]
 \includegraphics[width=1.0\textwidth]{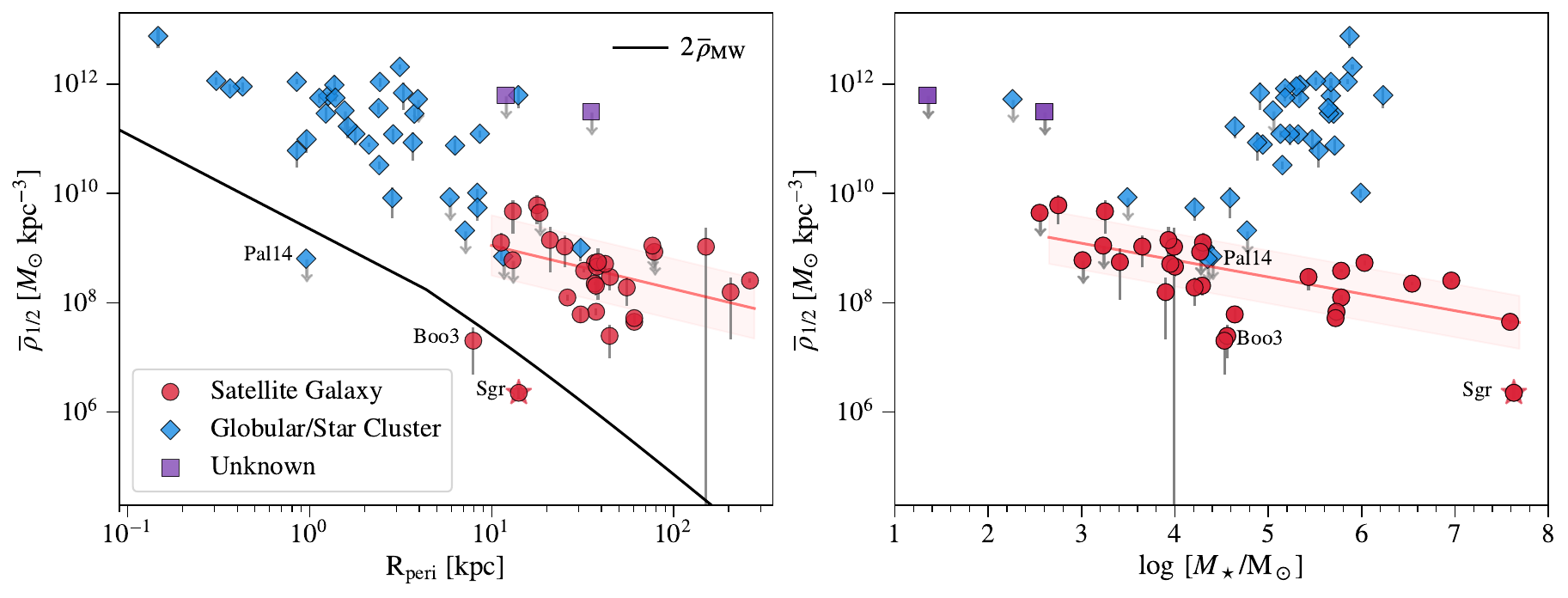}
\caption{ {\it Left:\/} The average enclosed mass density ($\overline{\rho}_{1/2}$) plotted against orbital pericenter (R$_{\rm peri}$).  The black line represents twice the enclosed Milky Way density as a function of radius (2$\overline{\rho}_{\rm MW}$).   Systems near this line are expected to show signs of tidal disruption.   While most systems are far from this threshold, we label systems close to the disruption line. {\it Right:}\/ We plot $\overline{\rho}_{1/2}$ against total stellar mass.  For satellite galaxies, lower luminosity systems tend to have higher enclosed mass density. \label{fig_density}}
\end{figure*}

\subsection{Enclosed Density and Tidal Influence on Milky Way Satellites}\label{ssec:tides}

The analysis above implicitly assumes that Milky Way satellites are in dynamical equilibrium.   We next examine orbital parameters to assess the degree to which Milky Way tidal forces may be affecting our measured quantities.   All of our satellite systems have proper motions measured by {\it Gaia} DR3, although in many cases this is based on very few stars.   We determine orbital parameters based on the proper motion and distance as compiled by \citet{pace2024}, and the DEIMOS radial velocity.  We use the {\tt galpy} orbit integration software \citep{bovy2015} and {\tt MWPotential2014} for the gravitational potential of the Milky Way, which is static in time and does not include the impact of the LMC.   As suggested by \citet{bovy2015}, we adopt a 50\% larger dark matter halo mass ($1.2\times10^{12}\Msun$) than the {\tt MWPotential2014} default.

To assess whether a system is vulnerable to gravitational disruption, we follow \citet{pace2022} by comparing the average enclosed density ($\overline{\rho}_{1/2}$) of each system to twice the enclosed Milky Way density ($\overline{\rho}_{\rm MW}$) at the same pericentric distance.  If a satellite sits below this line, its Jacobi tidal radius is larger than the half-light radius and it will likely be tidal disrupting.  In Figure~\ref{fig_density} (right panel), the majority of Milky Way satellite galaxies and globular clusters lie well above this threshold.  We label two satellite galaxies which sit close to this threshold:  the disrupting satellite galaxy Sgr dSph \citep{Majewski2003}, and Bo\"{o}tes~III which has long been suspected to be tidally disrupting \citep{Carlin2018, pace2022}. 

One globular cluster in Figure~\ref{fig_density} lies near the tidal-disruption threshold.  Palomar~14 is an outer halo system, and is previously known to have a low stellar density \citep{2011MNRAS.411.1989Z}.   While it is currently at a heliocentric distance of 74\,kpc and near apocenter, its orbital properties suggest a plunging orbit, with pericenter of less than 1\,kpc \citep{Baumgardt2019, Zonoozi2024}. 

We first focus on globular clusters as a population in Figure~\ref{fig_density}.   Globular clusters with the highest stellar densities ($\overline{\rho}_{1/2}$) are also the most massive systems with the smallest pericentric radii.   Globular clusters with lower stellar densities tend to have lower stellar mass and orbit in the outer Galactic halo.   Splitting the globular cluster sample at $\overline{\rho}_{1/2} = 10^{10}M_{\odot}$\,kpc$^{-3}$, lower density clusters have a mean current distance (pericenter) of 37\,kpc (14\,kpc),  compared to 16\,kpc (3\,kpc) for higher density globular clusters.   This can be understood as lower density clusters will not survive close passage near the Galactic plane, while more massive globular clusters are preferentially removed from the outer halo via dynamical friction \citep[e.g.,][]{gnedin1997}.

We next examine trends for satellite galaxies in Figure~\ref{fig_density}.  The average mass density, $\overline{\rho}_{1/2}$, mildly increases with both decreasing pericentric distance and decreasing stellar mass.   We again fit a linear model to  satellite galaxies only, first removing the two disrupting satellites and satellites in which the velocity dispersion is unresolved.  We report the coefficients of these fits in the Appendix. \citet{Kaplinghat2019} first pointed out a trend of increasing $\overline{\rho}$, enclosed within a fixed 150\,pc, with decreasing pericenter for classical Milky Way satellites, suggesting that only dense satellite galaxies can survive small pericenters.   While we see a similar trend for $\overline{\rho}_{1/2}$ versus $R_{\rm peri}$, the slope is just 2-$\sigma$ from a flat relation.   Interestingly, \citet[see their Figure 5;][]{pace2022} sees significantly more scatter between $\overline{\rho}_{1/2}$ and $R_{\rm peri}$, regardless of whether the LMC is included in the orbit integration.    Finally, it is curious that the minimum $\overline{\rho}_{1/2}$ values for both the satellite galaxy and globular cluster populations follow one magnitude above the Milky Way tidal threshold.

In the right panel of Figure~\ref{fig_density}, the relationship for satellite galaxies between $\overline{\rho}_{1/2}$ and stellar mass is perhaps easier to interpret.   As noted by \citet{Kravtsov2023}, if galaxies of larger stellar mass form in higher-mass
Cold Dark Matter (CDM) halos on average, but $r_{\rm 1/2}$ is roughly a fixed fraction of the virial radius, then larger stellar masses systems are predicted to have lower $\overline{\rho}_{1/2}$.   The observed trend suggests that lower stellar mass satellites, on average, are hosted in lower mass dark matter halos even at the extremely low stellar masses.  However, the large number of upper limits in density/velocity dispersion at the lowest stellar mass end may reduce or erase this trend when improved measurements are available.  We discuss this trend further in \S\,\ref{ssec:DM}.

\section{Metallicity}\label{sec_metals}

We next investigate the mean metallicity ($\overline{\rm [Fe/H]}$) and internal metallicity spread ($\sigma_{\rm [Fe/H]}$) in our Milky Way satellite sample.  In Paper I, we measured the combined equivalent widths of the Ca II triplet (CaT) lines (8498.0, 8542.1 and 8662.1\,\mbox{\AA}) and determined stellar metallicities, [Fe/H], for individual stars based on the  CaT-[Fe/H] calibration of 
\citet{Navabi2025}.  This calibration is an update to \citet{Carrera2013}, using the same functional form, with improved agreement with full spectrum-fitting results at the higher metallicity end ([Fe/H]$>-1.5$).   This empirical calibration is valid only for red giant branch stars, $M_V < 3$ (see Paper~I, Figure~13), and we thus remove horizontal branch stars and intrinsically fainter stars for this analysis.   These CaT-based [Fe/H] allow for homogeneous metallicity estimates across a wide range of signal-to-noise.  We again enforce a minimum number of 10 stars per system when determining metallicity spreads (55 systems), but relax this criteria to a minimum of 7 stars (61 systems) when determining the mean metallicity.  

To determine the mean metallicity, $\overline{\rm [Fe/H]}$, and internal metallicity spread, $\sigma_{\rm [Fe/H]}$, in each system, we assume the same two-parameter Gaussian model as detailed in \S\,\ref{ssec_vdisp}, in which the individual metallicity measurements and associated errors are described by a mean metallicity and internal metallicity spread.  We also use the same sampling algorithm, assuming flat priors, and test if the metallicity spread is resolved (non-zero) if the log of the ratio of Bayesian evidences is larger than unity.   Metallicities are again determined inside of 2\,$r_{\rm 1/2}$ to maintain uniformity, but are extended for globular clusters to a maximum of 4\,$r_{\rm 1/2}$ in cases where there are less than 10 available member stars.  We provide $\overline{\rm [Fe/H]}$, $\sigma_{\rm [Fe/H]}$, and associated errors for individual Milky Way systems in Table~\ref{table_objects}.

\begin{figure*}[t!]
 \includegraphics[width=1.0\textwidth]{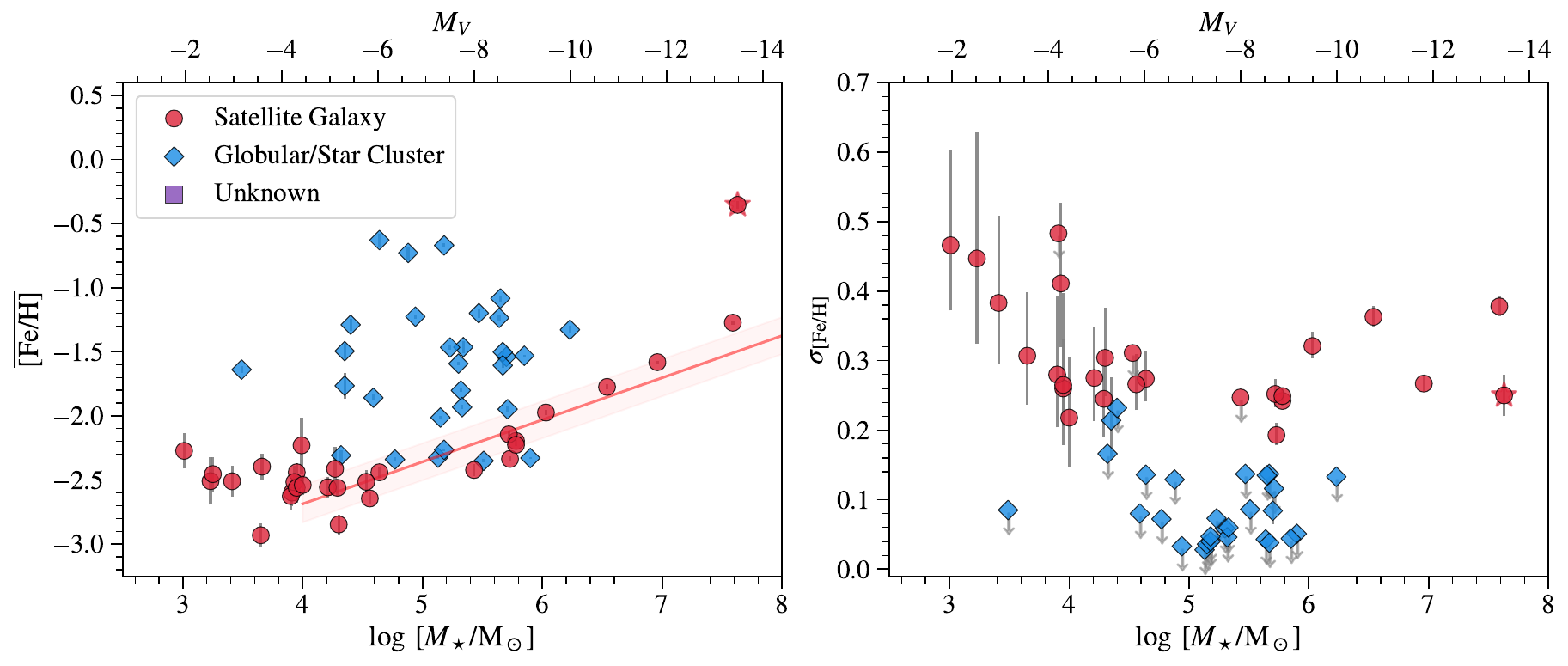}
\caption{{\it Left:\/} The mean stellar metallicity ($\overline{\rm [Fe/H]}$) and ({\it right}) the internal metallicity spread ($\sigma_{\rm [Fe/H]}$) versus total stellar mass.   Systems are color-coded as previous figures.  Satellite galaxies show the well-known mass-metallicity relationship down to $\log\;[M_\star/{\rm M}_\odot] \approx 4$.  Below this stellar masses threshold, the mass-metallicity relationship for satellite galaxies flattens and/or increases scatter.  The internal metallicity scatter shows a similar upturn or increased scatter at this stellar mass.  The red star symbol at $\log\;[M_\star/{\rm M}_\odot] \approx 7.6$ is the disrupting Sagittarius dSph which we do not included in our fits.    In contrast, globular clusters show no mass-metallicity relation and have internal metallicity dispersions below 0.2\,dex.  \label{fig_feh}}
\end{figure*}

\subsection{The Stellar Mass-Metallicity Relation}

The stellar mass-metallicity relationship is a fundamental scaling relation in galaxy evolution \citep[e.g.,][]{Gallazzi2005,Kirby2011, simon2019,bose2025}.   The slope and scatter of this relation provides insights into the balance between pristine gas inflow and metal-enriched outflows driven by stellar feedback. The tight observed scatter in this relationship ($\sim0.2$\,dex) has historically been used to argue against significant tidal disruption for Milky Way satellites \citep{simon07a,kirby2013}, however, \citet{riley2025}  demonstrated that significant tidal disruption may not increase scatter in the mass-metallicity relationship, but rather lead to a shallower slope.

In the left panel of Figure~\ref{fig_feh}, we plot stellar metallicity versus stellar mass ($\overline{\rm [Fe/H]}$ vs.~M$_\star$).  We first focus on globular clusters which show no trend in this panel.   Our DEIMOS Milky Way globular cluster sample shows the well-known apparent metallicity floor \citep{Harris2010}, such that no clusters lie below $\overline{\rm [Fe/H]} < -2.36$.  In the larger Milky Way sample of globular clusters, only one system is known at slightly lower metallicity, ESO 280-SC06, at $\overline{\rm [Fe/H]} < -2.54$ \citep{usman2025}.   In stark contrast, eight of our satellite galaxies have $\overline{\rm [Fe/H]} <-2.5$, and the majority of satellite galaxies are more metal-poor than our most metal-poor globular cluster, NGC\,6341 (M\,92) with $\overline{\rm [Fe/H]} = -2.35\pm0.02$.  The most metal-poor satellite galaxies in our sample is Eri\,IV with $\overline{\rm [Fe/H]} = -2.93\pm0.09$, consistent with \citet{heiger2024}.  The most metal-poor Milky Way satellite galaxy currently published is Pictor\,II with $\overline{\rm [Fe/H]} = -2.99\pm0.06$ \citep{pace2025}.

Satellite galaxies show a tight trend of decreasing metallicity with stellar mass down to $\log M_\star/M_\odot = 4$ ($M_V \sim -4.5$).   We first fit a linear model to satellite galaxies between $ 4 < \log M_\star/M_\odot < 8$ (16 systems), excluding the currently disrupting Sgr dSph (red star symbol in Figure~\ref{fig_feh}).  As described and reported in the Appendix, we find a fit of $\overline{\rm [Fe/H]} = 0.33^{+0.04}_{-0.04} \log[M_\star] -3.98^{+0.22}_{-0.21}$.  The average scatter in the intercept of this fit is 0.22\,dex.   We also compute scatter as the root-mean-square, finding a scatter of 0.16\,dex above $\log M_\star/M_\odot > 4$.  Compared to \citet{kirby2013} and \citet{simon2019}, our slope is slightly steeper ($0.33\pm{0.04}$ vs.~$0.29\pm0.02$) with similar scatter.  See the Appendix for further comparison to previous results. 

The behavior of the mass-metallicity relationship changes significantly at $\log M_\star/M_\odot \approx 4$ ($M_V \sim -4.5$).  Below this stellar mass, the trend becomes more shallow and/or increases in scatter (Figure~\ref{fig_feh}, left panel).  A flattening of the mass-metallicity relationship for extremely low stellar mass galaxies has been previously noted.   However, \citet{Fu2023} suggest a higher stellar mass transition of $\log M_\star/M_\odot \sim 5$ ($M_V > -7.0$) based on narrow-band photometric {\it HST} metallicities, while previous spectroscopic measurements suggest only an increase of scatter below $\log M_\star/M_\odot \sim 4$ \citep{simon2019,pace2024}.   Again, see the Appendix for a comparison to previous results. 

A stellar mass threshold below which the mass-metallicity relation changes is also seen in theoretical models of galaxy formation, but its value and interpretation remain uncertain. Earlier cosmological baryonic zoom-in simulations of isolated low-mass galaxies struggled to systematically recover the level of metal enrichment for galaxies with $\log M_\star/M_\odot \lesssim 5$, finding that these systems are especially sensitive to choices of star formation and stellar feedback prescriptions, nucleosynthetic yields, and Population III modeling \citep[e.g.,][]{maccio2017,wheeler2019,agetz2020,gandhi2022}.   However, both recent cosmological zoom-in simulations and semi-analytic models of low-mass galaxies in Milky Way-like environments show metal enrichment closer to observed Milky Way satellites, and suggest a flattening in the mass-metallicity relation between $\log M_\star/M_\odot < 4-5$ is due to either Population III IGM enrichment and enrichment from the host MW-mass system \citep{ahvazi2025, rey2025m} or choices of feedback prescriptions that affect galactic winds and metal loss/retention from the low-mass galaxies \citep{Manwadkar2022, rey2025}, although stochastic sampling may also play a role \citep{go2025, Andersson2025}.   See the Appendix for a direct comparison to selected predictions.

\subsection{Metallicity Spreads}\label{ssec:metal_spreads}

The presence of an internal metallicity spread is a key test in discerning between star clusters and galaxies \citep{willman2012}.   The underlying assumption is that dark matter-dominated galaxies better retain enriched ejecta from stellar feedback and form further generations of stars, increasing the metallicity differences between individual stars in a given system.   While globular clusters are now recognized to also host significant internal abundance spreads for light elements \citep{bastian2018}, the spread in overall metallicity is significantly smaller as compared to galaxies, with spectroscopic studies suggesting of order 
$\sigma_{\rm [Fe/H]} \sim 0.05$\,dex \citep{Carretta2009,Latour2025}.  

In the right panel of Figure~\ref{fig_feh}, we plot the internal metallicity spread versus stellar mass ($\sigma_{\rm [Fe/H]}$ vs.~M$_{\star}$). Nearly all globular clusters show unresolved metallicity spreads with a median 95\% upper limit of $\sigma_{\rm [Fe/H]} < 0.07$\,dex.  We do not plot one globular cluster which has a significant resolved metallicity dispersions: NGC\,2419.   NGC\,2419 has $\sigma_{\rm [Fe/H]}=0.19$\,dex, 
consistent with the prior CaT measurement by \citet{Cohen2010}.  However, \citet{larsen2019} interprets the CaT spread as an indicator of a spread in alpha-elements rather than iron; the true iron abundance dispersion of NGC\,2419 is likely much smaller \citep{bailin2019}.

In  contrast to globular clusters, the internal metallicity dispersion of satellite galaxies are 0.2\,dex and larger.    We determine the metallicity spread within 2 half-light radius and do not remove metallicity gradients which are present for the most luminous satellite galaxies ($\log M_\star/M_\odot > 6$).  For bright systems such as the Fornax and Sgr dSph,  this choice increases our values over the literature (see Appendix),  however, we believe this is a more fair comparison when including the faintest Milky Way systems for which there are too few stars to measure gradients. The median metallicity spread of the full Milky Way DEIMOS satellite galaxy sample is $\sigma_{\rm [Fe/H]} = 0.27$\,dex.  The scatter in $\sigma_{\rm [Fe/H]}$ is significant with systems ranging from 0.2-0.5\,dex.  The range in scatter is smaller than both previous measurements (see Appendix) and predictions from cosmological zoom-in simulations \citep{Escala2018,taylor2025,go2025}.   At the lowest stellar mass ($\log M_\star/M_\odot < 3.8$), $\sigma_{\rm [Fe/H]}$ appears to increase, however, these systems have large errors due to a small sample size (15 or fewer stars).

\begin{figure}[t!]
 \includegraphics[width=1.03\columnwidth]{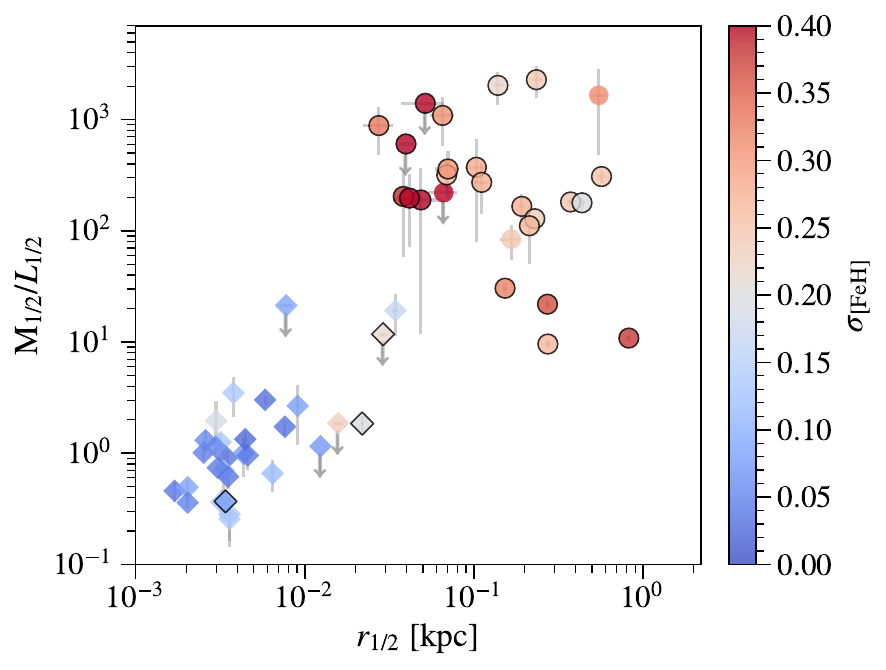}
\caption{Combining kinematic and metallicity measurements, we plot the mass-to-light ratio (M$_{\rm 1/2} / L_{1/2}$) versus the half-light radius ($r_{\rm 1/2}$).   Symbols are color-coded based on their internal metallicity dispersion ($\sigma_{\rm [Fe/H]}$). Systems with resolved metallicity dispersions are shown with black outlines.   Satellite galaxies (circles) and star clusters (diamonds) are well separated in these properties.\label{fig_sep}}
\end{figure}

\begin{figure*}[t!]
 \includegraphics[width=1.0\textwidth]{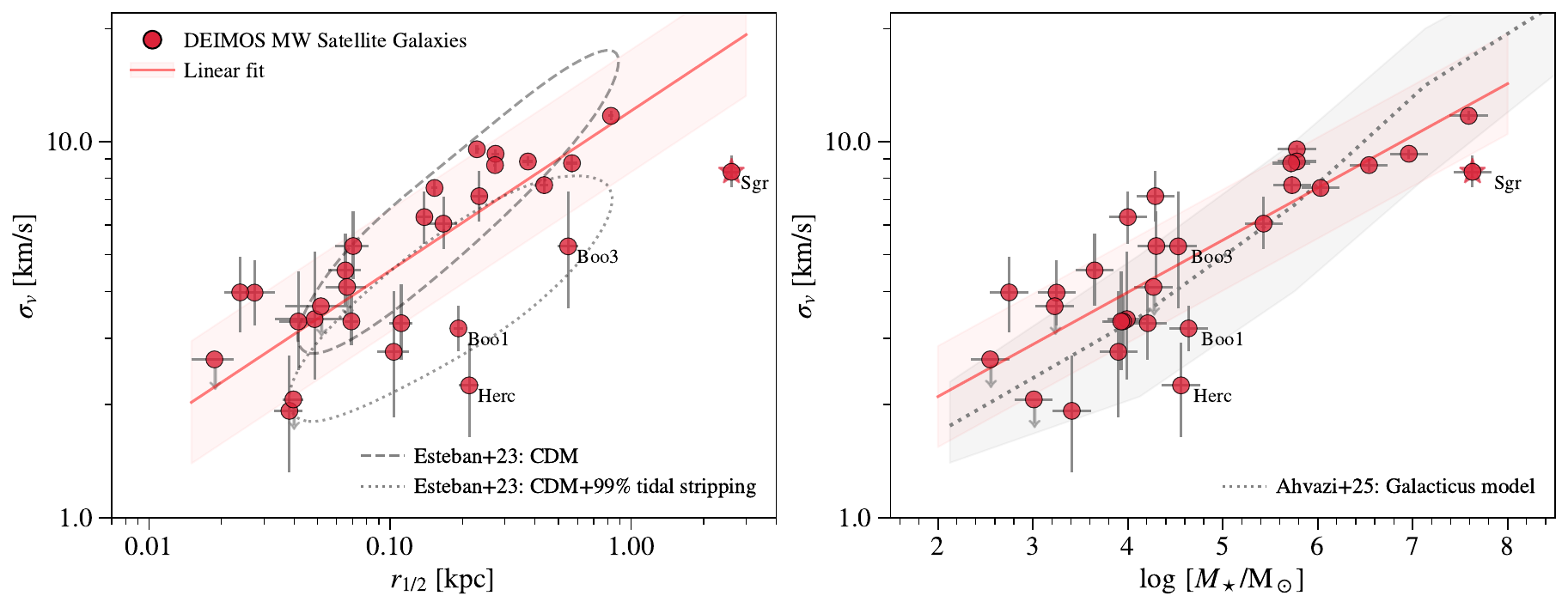}
\caption{{\it Left:}\/  The velocity dispersion versus half-light radius ($\sigma_v$ vs. $r_{\rm 1/2}$) for the DEIMOS satellite galaxies only.  The red line is a linear fit to the data, exclude the disrupting Sgr system (star symbol).   The dashed region is the 68\% enclosed region for a representative population of Milky Way-like satellites residing in CDM halos \citep{esteban2024}.   The dotted region is the same model with 99\% tidal stripping. {\it Right:}\/ The velocity dispersion plotted versus stellar mass ($\sigma_v$ vs. M$_{\star}$).   This agrees well with prediction from the semi-analytic model from \citet{ahvazi2025}.    Values for the linear fits (red lines) are provided in the Appendix. \label{fig_dwarfs}}
\end{figure*}

\section{Discussion}\label{sec:discussion}

We combine our kinematic and metallicity measurements, with the goal of better separating satellite galaxies from star clusters (\S\,\ref{ssec:classification}), and then consider satellite galaxies as probes of dark matter (\S\,\ref{ssec:DM}).

\subsection{Separating Galaxies and Star Clusters}\label{ssec:classification}

Cleanly separating satellite galaxies from globular clusters is particularly challenging for the Milky Way's faintest systems ($M_V > -4$).   Yet a clean census is critical to understanding how both populations formed \citep[e.g.,][]{Baumgardt2022, simon2024,smith2024,cerny2025}.  Given limited and heterogeneous data, there is no one way to classify systems.  Various indicators, or combination of indicators, are currently used including $M_V$, $r_{\rm 1/2}$, $\sigma_{\rm [Fe/H]}$, and stellar mass segregation.

In Figure~\ref{fig_sep}, we combine kinematic and photometric properties, plotting the mass-to-light ratio versus half-light radius ($M_{1/2}/L_{1/2}$ vs.~$r_{\rm 1/2}$), color-coding by the internal metallicity spread ($\sigma_{\rm [Fe/H]}$).   Systems in which the metallicity spread is resolved are outlined in black.   While this figure is not optimized for classification, it serves as a relatively clean separation between globular clusters and satellite galaxies.   We have already commented on one system, Sgr\,II, which does not separate fully (the blue diamond with $r_{\rm 1/2}=0.03$\,kpc).  Figure~\ref{fig_sep} demonstrates that while size alone remains a useful indicator of classification, there is overlap in the region between $r_{\rm 1/2} = 25 - 40$\,pc, and the addition of spectroscopic measurements (e.g., mass-to-light ratio and metallicity spread) significantly improves separation.  Nonetheless, recent predictions suggest a population of extremely low mass galaxies which would overlap even in this space with star clusters \citep[e.g.,][]{errani2024,taylor2025}.

\subsection{Satellite Galaxies:  Dark Matter tests}\label{ssec:DM}

The observed properties of extremely low mass galaxies provide strong constraint on the nature of dark matter \citep[e.g.,][]{bullock2017,sales2022,cruz2025, may2025,delos2025}. Rather than constrain any particular dark matter model, here we aim to provide robust observed relationships which can be used to constrain a wide range of dark matter model predictions.

In the left panel of Figure~\ref{fig_dwarfs}, we plot our DEIMOS-derived velocity dispersions versus half-light radius ($\sigma_v$ vs.~$r_{\rm 1/2}$) for satellite galaxies only.   Under the assumption that the each system represents a dark matter halo in CDM, the distribution of these points constrains the range of total halo mass and concentration for the Milky Way satellite galaxy population.  For example, \citet{esteban2024} showed that the scatter in $\sigma_v$ vs.~$r_{\rm 1/2}$ is consistent with expectations from CDM (dashed region, left panel Figure~\ref{fig_dwarfs}), with outliers to larger sizes and lower velocity dispersions being explained by tidal stripping (dotted region), similar to results from \citet{errani2022}.   We note that the observed $\sigma_v$ vs.~$r_{\rm 1/2}$ outliers in Figure~\ref{fig_dwarfs}, include the Hercules, Bo\"otes\,I and Bo\"otes\,III dSphs, all of which are suspected of tidal disruption \citep{ou2024, sandford2025, Carlin2018}.    We again fit a linear model to these data, providing the fit parameters in the Appendix.

In the right panel of Figure~\ref{fig_dwarfs}, we plot velocity dispersion against stellar mass ($\sigma_v$ vs.~M$_{\star}$).  While these data are roughly linear, there is significant scatter, particularly below 
$\log M_\star /M_\odot < 4.5$.   The distribution is marginally consistent with the semi-analytic predictions from \citet{ahvazi2025} in both slope and scatter.    Combining $\sigma_v$ and $r_{\rm 1/2}$, we plot the enclosed mass density, $\overline{\rho}_{1/2}$, in the right panel of Figure~\ref{fig_density}.  As discussed in \S\,\ref{ssec:tides}, the overall trend of increasing mass density with decreasing stellar mass is understood if galaxies of larger stellar mass form in higher-mass halos on average, and the fact that the half-light radius is a fixed fraction of the virial radius with some scatter \citep{Kravtsov2013,Kravtsov2023}.   

Finally, the Milky Way satellites remain attractive sites to search for signals of dark matter annihilation and decay at gamma-ray wavelengths \citep[e.g.,][]{Abdo2010,Strigari2018}. The strength of these signals depend on details of the dark matter particle itself and the combined central dark matter density and heliocentric distance of a given satellite galaxy. The latter is often parameterized by the `J-factor'.   Using the scaling relations from \citet{pace2019}, our highest J-factor systems is Segue\,1, $\log_{10}J\,[{\rm GeV}^2\,{\rm cm}^{-5}] = 19.4\pm0.4$.  This object is also among the highest ranked systems in \citet{mcdaniel2024}, however, because J-factors scale as velocity dispersion to the fourth power, our reduced velocity dispersion errors translates into smaller J-factor errors.  Meaningful constraints on the dark matter particle are currently determined from stacked analysis of all Milky Way satellite galaxies \citep[e.g.][]{mcdaniel2024}, again emphasizing the need for uniformly determined spectroscopic quantities.    We leave a combined analyses based on this uniform DEIMOS sample to a future contribution.

\section{Summary}\label{sec:summary}

We present the largest self-consistent sample of spectroscopically-derived quantities
for Milky Way satellite galaxies and globular clusters.   The sample is based on a homogeneous re-analysis of stars observed with the Keck/DEIMOS spectrograph \citep[Paper I,][]{geha_paper1}.   We determine uniform internal velocity dispersions, enclosed dynamical masses, mean [Fe/H] metallicities and internal metallicity spreads ($\sigma_{\rm [Fe/H]}$) for systems with 10 or more member stars, and examine various scaling relations for these populations.   

Previous population studies of the Milky Way's stellar satellites have largely, and necessarily, relied on heterogeneous datasets when examining spectroscopic-based properties. While in many cases our results agree with the literature, our uniform analysis reduces the possibility of systematics due to aggregated data compilations.  In summary, we find:

\begin{itemize}

 \item At a given stellar mass, systems classified as satellite galaxies are well separated from globular clusters in their dynamical mass and mass-to-light ratios. 
 
 \item For satellite galaxies, mass-to-light ratio increases with decreasing stellar mass.   The maximum observed value is M$_{\rm 1/2} / L_{1/2}\sim 2500 \pm 1200$ for Segue\,1 ($\log M_{\star}/M_\odot = 2.75$).  
 
 \item For globular clusters, the enclosed stellar density ($\overline{\rho}_{1/2}$) is largest for globular clusters with small orbital pericenters and large stellar masses,  consistent with dynamical processes acting on the Galactic globular cluster population.
 
 \item For globular clusters, we recover the known metallicity `floor', with no observed clusters below $\overline{\rm [Fe/H]} < -2.4$.

\item For satellite galaxies, the stellar-mass metallicity relationship is similar to previous work, but with a slightly steeper slope:  $\overline{\rm [Fe/H]} = 0.33^{+0.04}_{-0.04} \log[M_\star] -3.98^{+0.22}_{-0.21}$.   The rms scatter around this relationship is 0.16\,dex.

 \item For satellite galaxies below $\log M_{\star}/M_\odot \approx 4$, the mass-metallicity relation flattens and/or increases in scatter. 
 
 \item  Satellite galaxies have internal metallicity scatter that is constant between 0.3-0.4\,dex across a wide range of stellar mass.

\item For satellite galaxies, the distribution of $\sigma_v$ vs.~$r_{\rm 1/2}$ and $\overline{\rho}_{1/2}$ vs.$\log M_{\star}$ is consistent with CDM predictions, with a handful of outliers explained by tidal disruption.  

\item We provide measured quantities for all 67 systems presented in this work.  This includes velocity dispersions, enclosed dynamical masses, $\overline{\rm [Fe/H]}$, $\sigma_{\rm [Fe/H]}$, and associated errors (Table~\ref{table_objects}).

\end{itemize}

Due to proximity, the Milky Way's population of satellite galaxies and star clusters includes, and will continue to include, the lowest mass and least chemically enriched stellar systems in the known Universe.   The number of satellites, particularly at the faintest luminosities, is expected to increase dramatically with highly anticipated upcoming wide-field imaging surveys including LSST \citep{lsst2019}, the Roman Space Telescope \citep{roman2015} and Euclid mission \citep{euclid2025}.  

Our work highlights the importance of homogeneous spectroscopic follow-up of individual stars 
and a uniform analysis of properties in both known and anticipated low mass stellar systems around the Milky Way and beyond.   In a future contribution, we will present a uniform analysis of Keck/DEIMOS archival data of M\,31 satellites, and work towards homogeneously merging other existing spectroscopic datasets \citep[e.g.,][]{Tsantaki2022}.   Careful analysis of current spectroscopic datasets can provide key first-epoch spectroscopic measurements for current and future highly multiplexed spectroscopic surveys targeting individual stars in Milky Way satellites such as  DESI \citep{Cooper2023}, PFS \citep{pfs2024}, 4MOST \citep{4MOST2023}, and the \href{http:/via-project.org}{Via Project}.  Such work will ultimately enable tighter constraints on the low mass stellar mass-halo mass relation, improved J-factor estimates for dark matter searches, and lay a foundation for interpreting the flood of new Milky Way satellites expected in the LSST/Roman era.
 
\begin{acknowledgments} 
MG was supported in part by a grant~from the Howard Hughes Medical Institute (HHMI) through the HHMI Professors Program.  We thank Yasmeen Asali, William Cerny, Vedant Chandra, Ivan Esteban, Erin Kado-Fong, Pratik Gandhi, Viraj Manwadkar, Sebastian Monzon, Ethan Nadler, Annika Peter, Frank van den Bosch and Risa Wechsler for insightful comments that improved this work.    This work has made use of the Local Volume Database \citep{pace2024}.  This research has made extensive use of the Keck Observatory Archive (KOA), which is operated by the W. M. Keck Observatory and the NASA Exoplanet Science Institute (NExScI), under contract with the National Aeronautics and Space Administration.  

\end{acknowledgments}

\software{This research made use of many community-developed or community-maintained software packages, including (in alphabetical order):
Astropy \citep{astropy},
galpy \citep{bovy2014}
IPython \citep{ipython},
Matplotlib \citep{matplotlib},
NumPy \citep{numpy},
Emcee \citep{emcee},
and SciPy \citep{scipy}.
This research has also made use of NASA's Astrophysics Data System.
}

\facility{Keck: DEIMOS}

\bibliographystyle{yahapj}
\bibliography{bib_deimos,software}

@ARTICLE{sandford2025,
       author = {{Sandford}, Nathan R. and {Li}, Ting S. and {Koposov}, Sergey E. and {Hayashi}, Kohei and {Pace}, Andrew B. and {Erkal}, Denis and {Bovy}, Jo and {Da Costa}, Gary S. and {Cullinane}, Lara R. and {Ji}, Alexander P. and et al.},
        title = "{Chemodynamics of Bo{\"o}tesI with $S^{5}$: Revised Velocity Gradient, Dark Matter Density, and Galactic Chemical Evolution Constraints}",
      journal = {arXiv e-prints},
         year = 2025,
        month = sep,
          eid = {arXiv:2509.02546},
        pages = {arXiv:2509.02546},
          doi = {10.48550/arXiv.2509.02546},
archivePrefix = {arXiv},
       eprint = {2509.02546},
 primaryClass = {astro-ph.GA},
       adsurl = {https://ui.adsabs.harvard.edu/abs/2025arXiv250902546S}
}

@ARTICLE{cruz2025,
       author = {{Cruz}, Akaxia and {Brooks}, Alyson and {Lisanti}, Mariangela and {Peter}, Annika H.~G. and {Geda}, Robel and {Quinn}, Thomas and {Tremmel}, Michael and {Munshi}, Ferah and {Keller}, Ben and {Wadsley}, James},
        title = "{Dwarf diversity in $Λ$CDM with baryons}",
      journal = {arXiv e-prints},
         year = 2025,
        month = oct,
          eid = {arXiv:2510.11800},
        pages = {arXiv:2510.11800},
          doi = {10.48550/arXiv.2510.11800},
archivePrefix = {arXiv},
       eprint = {2510.11800},
 primaryClass = {astro-ph.GA},
       adsurl = {https://ui.adsabs.harvard.edu/abs/2025arXiv251011800C}
}

@ARTICLE{sales2022,
       author = {{Sales}, Laura V. and {Wetzel}, Andrew and {Fattahi}, Azadeh},
        title = "{Baryonic solutions and challenges for cosmological models of dwarf galaxies}",
      journal = {Nature Astronomy},
         year = 2022,
        month = jun,
       volume = {6},
        pages = {897-910},
          doi = {10.1038/s41550-022-01689-w},
archivePrefix = {arXiv},
       eprint = {2206.05295},
 primaryClass = {astro-ph.GA},
       adsurl = {https://ui.adsabs.harvard.edu/abs/2022NatAs...6..897S}
}

@ARTICLE{Escala2018,
       author = {{Escala}, Ivanna and {Wetzel}, Andrew and {Kirby}, Evan N. and {Hopkins}, Philip F. and {Ma}, Xiangcheng and {Wheeler}, Coral and {Kere{\v{s}}}, Du{\v{s}}an and {Faucher-Gigu{\`e}re}, Claude-Andr{\'e} and {Quataert}, Eliot},
        title = "{Modelling chemical abundance distributions for dwarf galaxies in the Local Group: the impact of turbulent metal diffusion}",
      journal = {\mnras},
         year = 2018,
        month = feb,
       volume = {474},
       number = {2},
        pages = {2194-2211},
          doi = {10.1093/mnras/stx2858},
archivePrefix = {arXiv},
       eprint = {1710.06533},
 primaryClass = {astro-ph.GA},
       adsurl = {https://ui.adsabs.harvard.edu/abs/2018MNRAS.474.2194E}
}

@ARTICLE{gandhi2022,
       author = {{Gandhi}, Pratik J. and {Wetzel}, Andrew and {Hopkins}, Philip F. and {Shappee}, Benjamin J. and {Wheeler}, Coral and {Faucher-Gigu{\`e}re}, Claude-Andr{\'e}},
        title = "{Exploring metallicity-dependent rates of Type Ia supernovae and their impact on galaxy formation}",
      journal = {\mnras},
         year = 2022,
        month = oct,
       volume = {516},
       number = {2},
        pages = {1941-1958},
          doi = {10.1093/mnras/stac2228},
archivePrefix = {arXiv},
       eprint = {2202.10477},
 primaryClass = {astro-ph.GA},
       adsurl = {https://ui.adsabs.harvard.edu/abs/2022MNRAS.516.1941G}
}

@ARTICLE{Gallazzi2005,
       author = {{Gallazzi}, Anna and {Charlot}, St{\'e}phane and {Brinchmann}, Jarle and {White}, Simon D.~M. and {Tremonti}, Christy A.},
        title = "{The ages and metallicities of galaxies in the local universe}",
      journal = {\mnras},
         year = 2005,
        month = sep,
       volume = {362},
       number = {1},
        pages = {41-58},
          doi = {10.1111/j.1365-2966.2005.09321.x},
archivePrefix = {arXiv},
       eprint = {astro-ph/0506539},
 primaryClass = {astro-ph},
       adsurl = {https://ui.adsabs.harvard.edu/abs/2005MNRAS.362...41G}
}

@ARTICLE{bose2025,
       author = {{Bose}, Sownak and {Deason}, Alis J.},
        title = "{A galactic tug-of-war: how (not) to simultaneously fit the Milky Way satellite luminosity function and the mass-metallicity relation}",
      journal = {arXiv e-prints},
         year = 2025,
        month = sep,
          eid = {arXiv:2509.07066},
        pages = {arXiv:2509.07066},
          doi = {10.48550/arXiv.2509.07066},
archivePrefix = {arXiv},
       eprint = {2509.07066},
 primaryClass = {astro-ph.GA},
       adsurl = {https://ui.adsabs.harvard.edu/abs/2025arXiv250907066B}
}

@ARTICLE{go2025,
       author = {{Go}, Minsung and {Jeon}, Myoungwon and {Choi}, Yumi and {Kallivayalil}, Nitya and {Sohn}, Sangmo Tony and {Besla}, Gurtina and {Richstein}, Hannah and {Fu}, Sal Wanying and {Jeong}, Tae Bong and {Shin}, Jihye},
        title = "{Understanding Stellar Mass{\textendash}Metallicity and Size Relations in Simulated Ultrafaint Dwarf Galaxies}",
      journal = {\apj},
         year = 2025,
        month = jun,
       volume = {986},
       number = {2},
          eid = {214},
        pages = {214},
          doi = {10.3847/1538-4357/add2fa},
archivePrefix = {arXiv},
       eprint = {2411.14683},
 primaryClass = {astro-ph.GA},
       adsurl = {https://ui.adsabs.harvard.edu/abs/2025ApJ...986..214G}
}

@ARTICLE{bovy2015,
       author = {{Bovy}, Jo},
        title = "{galpy: A python Library for Galactic Dynamics}",
      journal = {\apjs},
         year = 2015,
        month = feb,
       volume = {216},
       number = {2},
          eid = {29},
        pages = {29},
          doi = {10.1088/0067-0049/216/2/29},
archivePrefix = {arXiv},
       eprint = {1412.3451},
 primaryClass = {astro-ph.GA},
       adsurl = {https://ui.adsabs.harvard.edu/abs/2015ApJS..216...29B}
}

@ARTICLE{Battaglia2008,
       author = {{Battaglia}, G. and {Helmi}, A. and {Tolstoy}, E. and {Irwin}, M. and {Hill}, V. and {Jablonka}, P.},
        title = "{The Kinematic Status and Mass Content of the Sculptor Dwarf Spheroidal Galaxy}",
      journal = {\apjl},
         year = 2008,
        month = jul,
       volume = {681},
       number = {1},
        pages = {L13},
          doi = {10.1086/590179},
archivePrefix = {arXiv},
       eprint = {0802.4220},
 primaryClass = {astro-ph},
       adsurl = {https://ui.adsabs.harvard.edu/abs/2008ApJ...681L..13B}
}

@ARTICLE{geha_paper1,
       author = {{Geha}, Marla and {Pelliccia}, Debora and {Prochaska}, J. Xavier and {Cerny}, William and {Davies}, Frederick B. and {Hennawi}, Joseph and {Holden}, Brad and {Reichwein}, Dusty and {Westfall}, Kyle B.},
        title = "{The Keck/DEIMOS Stellar Archive: I. Uniform Velocities and Metallicities for 78 Milky Way Dwarf Galaxies and Globular Clusters}",
      journal = {\apj},
         year = 2026,
        pages = {https://doi.org/10.3847/1538-4357/ae290d},
       adsurl = {https://doi.org/10.3847/1538-4357/ae290d},
          doi = {10.3847/1538-4357/ae290d}
}

@ARTICLE{pfs2024,
       author = {{Hirai}, Yutaka and {Kirby}, Evan N. and {Chiba}, Masashi and {Hayashi}, Kohei and {Anguiano}, Borja and {Saitoh}, Takayuki R. and {Ishigaki}, Miho N. and {Beers}, Timothy C.},
        title = "{Chemo-dynamical Evolution of Simulated Satellites for a Milky Way{\textendash}like Galaxy}",
      journal = {\apj},
         year = 2024,
        month = aug,
       volume = {970},
       number = {2},
          eid = {105},
        pages = {105},
          doi = {10.3847/1538-4357/ad500c},
archivePrefix = {arXiv},
       eprint = {2405.05330},
 primaryClass = {astro-ph.GA},
       adsurl = {https://ui.adsabs.harvard.edu/abs/2024ApJ...970..105H}
}

@ARTICLE{daneli2023,
       author = {{Danieli}, Shany and {Greene}, Jenny E. and {Carlsten}, Scott and {Jiang}, Fangzhou and {Beaton}, Rachael and {Goulding}, Andy D.},
        title = "{ELVES. IV. The Satellite Stellar-to-halo Mass Relation Beyond the Milky Way}",
      journal = {\apj},
         year = 2023,
        month = oct,
       volume = {956},
       number = {1},
          eid = {6},
        pages = {6},
          doi = {10.3847/1538-4357/acefbd},
archivePrefix = {arXiv},
       eprint = {2210.14233},
 primaryClass = {astro-ph.GA},
       adsurl = {https://ui.adsabs.harvard.edu/abs/2023ApJ...956....6D}
}

@ARTICLE{gnedin1997,
       author = {{Gnedin}, Oleg Y. and {Ostriker}, Jeremiah P.},
        title = "{Destruction of the Galactic Globular Cluster System}",
      journal = {\apj},
         year = 1997,
        month = jan,
       volume = {474},
       number = {1},
        pages = {223-255},
          doi = {10.1086/303441},
archivePrefix = {arXiv},
       eprint = {astro-ph/9603042},
 primaryClass = {astro-ph},
       adsurl = {https://ui.adsabs.harvard.edu/abs/1997ApJ...474..223G}
}

@ARTICLE{Majewski2003,
       author = {{Majewski}, Steven R. and {Skrutskie}, M.~F. and {Weinberg}, Martin D. and {Ostheimer}, James C.},
        title = "{A Two Micron All Sky Survey View of the Sagittarius Dwarf Galaxy. I. Morphology of the Sagittarius Core and Tidal Arms}",
      journal = {\apj},
         year = 2003,
        month = dec,
       volume = {599},
       number = {2},
        pages = {1082-1115},
          doi = {10.1086/379504},
archivePrefix = {arXiv},
       eprint = {astro-ph/0304198},
 primaryClass = {astro-ph},
       adsurl = {https://ui.adsabs.harvard.edu/abs/2003ApJ...599.1082M}
}

@ARTICLE{wheeler2025,
       author = {{Wheeler}, Coral and {Moreno}, Jorge and {Rodriguez Wimberly}, M. Katy and {Mercado}, Francisco J. and {Bullock}, James S. and {Boylan-Kolchin}, Michael and {Gandhi}, Pratik J. and {Loebman}, Sarah R. and {Hopkins}, Philip F.},
        title = "{How invisible stellar halos bias our understanding of ultra-faint galaxies}",
      journal = {arXiv e-prints},
         year = 2025,
        month = jun,
          eid = {arXiv:2506.15785},
        pages = {arXiv:2506.15785},
          doi = {10.48550/arXiv.2506.15785},
archivePrefix = {arXiv},
       eprint = {2506.15785},
 primaryClass = {astro-ph.GA},
       adsurl = {https://ui.adsabs.harvard.edu/abs/2025arXiv250615785W}
}

@ARTICLE{2011MNRAS.411.1989Z,
       author = {{Zonoozi}, Akram Hasani and {K{\"u}pper}, Andreas H.~W. and {Baumgardt}, Holger and {Haghi}, Hosein and {Kroupa}, Pavel and {Hilker}, Michael},
        title = "{Direct N-body simulations of globular clusters - I. Palomar 14}",
      journal = {\mnras},
         year = 2011,
        month = mar,
       volume = {411},
       number = {3},
        pages = {1989-2001},
          doi = {10.1111/j.1365-2966.2010.17831.x},
archivePrefix = {arXiv},
       eprint = {1010.2210},
 primaryClass = {astro-ph.GA},
       adsurl = {https://ui.adsabs.harvard.edu/abs/2011MNRAS.411.1989Z}
}

@ARTICLE{kirby2013,
       author = {{Kirby}, Evan N. and {Cohen}, Judith G. and {Guhathakurta}, Puragra and {Cheng}, Lucy and {Bullock}, James S. and {Gallazzi}, Anna},
        title = "{The Universal Stellar Mass-Stellar Metallicity Relation for Dwarf Galaxies}",
      journal = {\apj},
         year = 2013,
        month = dec,
       volume = {779},
       number = {2},
          eid = {102},
        pages = {102},
          doi = {10.1088/0004-637X/779/2/102},
archivePrefix = {arXiv},
       eprint = {1310.0814},
 primaryClass = {astro-ph.GA},
       adsurl = {https://ui.adsabs.harvard.edu/abs/2013ApJ...779..102K}
}

@ARTICLE{rey2025m,
       author = {{Rey}, Martin P. and {Katz}, Harley and {Cadiou}, Corentin and {Sanati}, Mahsa and {Agertz}, Oscar and {Blaizot}, Jeremy and {Cameron}, Alex J. and {Choustikov}, Nicholas and {Devriendt}, Julien and {Hauk}, Uliana and et al.},
        title = "{MEGATRON: how the first stars create an iron metallicity plateau in the smallest dwarf galaxies}",
      journal = {arXiv e-prints},
         year = 2025,
        month = oct,
          eid = {arXiv:2510.05232},
        pages = {arXiv:2510.05232},
          doi = {10.48550/arXiv.2510.05232},
archivePrefix = {arXiv},
       eprint = {2510.05232},
 primaryClass = {astro-ph.GA},
       adsurl = {https://ui.adsabs.harvard.edu/abs/2025arXiv251005232R}
}

@ARTICLE{Baumgardt2022,
       author = {{Baumgardt}, H. and {Faller}, J. and {Meinhold}, N. and {McGovern-Greco}, C. and {Hilker}, M.},
        title = "{Stellar mass segregation as separating classifier between globular clusters and ultrafaint dwarf galaxies}",
      journal = {\mnras},
         year = 2022,
        month = mar,
       volume = {510},
       number = {3},
        pages = {3531-3545},
          doi = {10.1093/mnras/stab3629},
archivePrefix = {arXiv},
       eprint = {2112.04689},
 primaryClass = {astro-ph.GA},
       adsurl = {https://ui.adsabs.harvard.edu/abs/2022MNRAS.510.3531B}
}

@ARTICLE{esteban2024,
       author = {{Esteban}, Ivan and {Peter}, Annika H.~G. and {Kim}, Stacy Y.},
        title = "{Milky Way satellite velocities reveal the dark matter power spectrum at small scales}",
      journal = {\prd},
         year = 2024,
        month = dec,
       volume = {110},
       number = {12},
          eid = {123013},
        pages = {123013},
          doi = {10.1103/PhysRevD.110.123013},
archivePrefix = {arXiv},
       eprint = {2306.04674},
 primaryClass = {astro-ph.CO},
       adsurl = {https://ui.adsabs.harvard.edu/abs/2024PhRvD.110l3013E}
}

@ARTICLE{brown2025,
       author = {{Brown}, Shaun T. and {Fattahi}, Azadeh and {Gutcke}, Thales A. and {Ploeckinger}, Sylvia and {Sureda}, Joaquin and {Bose}, Sownak and {Doppel}, Jessica E. and {Pakmor}, R{\"u}diger and {Jenkins}, Adrian},
        title = "{LYRA ultra-faints: The emergence of faint dwarf galaxies in the presence of an early Lyman-Werner background}",
      journal = {arXiv e-prints},
         year = 2025,
        month = nov,
          eid = {arXiv:2511.21824},
        pages = {arXiv:2511.21824},
          doi = {10.48550/arXiv.2511.21824},
archivePrefix = {arXiv},
       eprint = {2511.21824},
 primaryClass = {astro-ph.GA},
       adsurl = {https://ui.adsabs.harvard.edu/abs/2025arXiv251121824B}
}

@ARTICLE{cerny2025,
       author = {{Cerny}, William and {Bissonette}, Daisy and {Ji}, Alexander P. and {Geha}, Marla and {Chiti}, Anirudh and {Smith}, Simon E.~T. and {Simon}, Joshua D. and {Pace}, Andrew B. and {Kirby}, Evan N. and {Venn}, Kim A. and et al.},
        title = "{No Observational Evidence for Dark Matter Nor a Large Metallicity Spread in the Extreme Milky Way Satellite Ursa Major III / UNIONS 1}",
      journal = {arXiv e-prints},
         year = 2025,
        month = oct,
          eid = {arXiv:2510.02431},
        pages = {arXiv:2510.02431},
          doi = {10.48550/arXiv.2510.02431},
archivePrefix = {arXiv},
       eprint = {2510.02431},
 primaryClass = {astro-ph.GA},
       adsurl = {https://ui.adsabs.harvard.edu/abs/2025arXiv251002431C}
}

@ARTICLE{maccio2017,
       author = {{Macci{\`o}}, Andrea V. and {Frings}, Jonas and {Buck}, Tobias and {Penzo}, Camilla and {Dutton}, Aaron A. and {Blank}, Marvin and {Obreja}, Aura},
        title = "{The edge of galaxy formation - I. Formation and evolution of MW-satellite analogues before accretion}",
      journal = {\mnras},
         year = 2017,
        month = dec,
       volume = {472},
       number = {2},
        pages = {2356-2366},
          doi = {10.1093/mnras/stx2048},
archivePrefix = {arXiv},
       eprint = {1707.01106},
 primaryClass = {astro-ph.GA},
       adsurl = {https://ui.adsabs.harvard.edu/abs/2017MNRAS.472.2356M}
}

@ARTICLE{agetz2020,
       author = {{Agertz}, Oscar and {Pontzen}, Andrew and {Read}, Justin I. and {Rey}, Martin P. and {Orkney}, Matthew and {Rosdahl}, Joakim and {Teyssier}, Romain and {Verbeke}, Robbert and {Kretschmer}, Michael and {Nickerson}, Sarah},
        title = "{EDGE: the mass-metallicity relation as a critical test of galaxy formation physics}",
      journal = {\mnras},
         year = 2020,
        month = jan,
       volume = {491},
       number = {2},
        pages = {1656-1672},
          doi = {10.1093/mnras/stz3053},
archivePrefix = {arXiv},
       eprint = {1904.02723},
 primaryClass = {astro-ph.GA},
       adsurl = {https://ui.adsabs.harvard.edu/abs/2020MNRAS.491.1656A}
}

@ARTICLE{wheeler2019,
       author = {{Wheeler}, Coral and {Hopkins}, Philip F. and {Pace}, Andrew B. and {Garrison-Kimmel}, Shea and {Boylan-Kolchin}, Michael and {Wetzel}, Andrew and {Bullock}, James S. and {Kere{\v{s}}}, Du{\v{s}}an and {Faucher-Gigu{\`e}re}, Claude-Andr{\'e} and {Quataert}, Eliot},
        title = "{Be it therefore resolved: cosmological simulations of dwarf galaxies with 30 solar mass resolution}",
      journal = {\mnras},
         year = 2019,
        month = dec,
       volume = {490},
       number = {3},
        pages = {4447-4463},
          doi = {10.1093/mnras/stz2887},
archivePrefix = {arXiv},
       eprint = {1812.02749},
 primaryClass = {astro-ph.GA},
       adsurl = {https://ui.adsabs.harvard.edu/abs/2019MNRAS.490.4447W}
}

@ARTICLE{Andersson2025,
       author = {{Andersson}, Eric P. and {Rey}, Martin P. and {Yates}, Robert M. and {Read}, Justin I. and {Agertz}, Oscar and {Ji}, Alexander P. and {Mead}, Jennifer and {Brauer}, Kaley and {Mac Low}, Mordecai-Mark},
        title = "{EDGE-INFERNO: How chemical enrichment assumptions impact the individual stars of a simulated ultra-faint dwarf galaxy}",
      journal = {arXiv e-prints},
         year = 2025,
        month = nov,
          eid = {arXiv:2511.05695},
        pages = {arXiv:2511.05695},
          doi = {10.48550/arXiv.2511.05695},
archivePrefix = {arXiv},
       eprint = {2511.05695},
 primaryClass = {astro-ph.GA},
       adsurl = {https://ui.adsabs.harvard.edu/abs/2025arXiv251105695A}
}

@ARTICLE{delos2025,
       author = {{Delos}, M. Sten and {Ahvazi}, Niusha and {Benson}, Andrew},
        title = "{Testing warm dark matter with kinematics of the smallest galaxies}",
      journal = {arXiv e-prints},
         year = 2025,
        month = dec,
          eid = {arXiv:2512.04156},
        pages = {arXiv:2512.04156},
          doi = {10.48550/arXiv.2512.04156},
archivePrefix = {arXiv},
       eprint = {2512.04156},
 primaryClass = {astro-ph.CO},
       adsurl = {https://ui.adsabs.harvard.edu/abs/2025arXiv251204156D}
}

@ARTICLE{nadler2020,
       author = {{Nadler}, E.~O. and {Wechsler}, R.~H. and {Bechtol}, K. and {Mao}, Y.-Y. and {Green}, G. and {Drlica-Wagner}, A. and {McNanna}, M. and {Mau}, S. and {Pace}, A.~B. and {Simon}, J.~D. and et al.},
        title = "{Milky Way Satellite Census. II. Galaxy-Halo Connection Constraints Including the Impact of the Large Magellanic Cloud}",
      journal = {\apj},
         year = 2020,
        month = apr,
       volume = {893},
       number = {1},
          eid = {48},
        pages = {48},
          doi = {10.3847/1538-4357/ab846a},
archivePrefix = {arXiv},
       eprint = {1912.03303},
 primaryClass = {astro-ph.GA},
       adsurl = {https://ui.adsabs.harvard.edu/abs/2020ApJ...893...48N}
}

@ARTICLE{yang2025,
       author = {{Yang}, Hao and {Wang}, Wenting and {Zhu}, Ling and {Li}, Ting S. and {Koposov}, Sergey E. and {Han}, Jiaxin and {Li}, Songting and {Shi}, Rui and {Valluri}, Monica and {Riley}, Alexander H. and et al.},
        title = "{The Dark Matter Content of Milky Way Dwarf Spheroidal Galaxies: Draco, Sextans, and Ursa Minor}",
      journal = {\apj},
         year = 2025,
        month = nov,
       volume = {993},
       number = {2},
          eid = {249},
        pages = {249},
          doi = {10.3847/1538-4357/ae07ce},
archivePrefix = {arXiv},
       eprint = {2507.02284},
 primaryClass = {astro-ph.GA},
       adsurl = {https://ui.adsabs.harvard.edu/abs/2025ApJ...993..249Y}
}

@ARTICLE{heiger2024,
       author = {{Heiger}, M.~E. and {Li}, T.~S. and {Pace}, A.~B. and {Simon}, J.~D. and {Ji}, A.~P. and {Chiti}, A. and {Bom}, C.~R. and {Carballo-Bello}, J.~A. and {Carlin}, J.~L. and {Cerny}, W. and et al.},
        title = "{Reading between the (Spectral) Lines: Magellan/IMACS Spectroscopy of the Ultrafaint Dwarf Galaxies Eridanus IV and Centaurus I}",
      journal = {\apj},
         year = 2024,
        month = feb,
       volume = {961},
       number = {2},
          eid = {234},
        pages = {234},
          doi = {10.3847/1538-4357/ad0cf7},
archivePrefix = {arXiv},
       eprint = {2308.08602},
 primaryClass = {astro-ph.GA},
       adsurl = {https://ui.adsabs.harvard.edu/abs/2024ApJ...961..234H}
}

@ARTICLE{battaglia2013,
       author = {{Battaglia}, Giuseppina and {Helmi}, Amina and {Breddels}, Maarten},
        title = "{Internal kinematics and dynamical models of dwarf spheroidal galaxies around the Milky Way}",
      journal = {\nar},
         year = 2013,
        month = sep,
       volume = {57},
       number = {3-4},
        pages = {52-79},
          doi = {10.1016/j.newar.2013.05.003},
archivePrefix = {arXiv},
       eprint = {1305.5965},
 primaryClass = {astro-ph.CO},
       adsurl = {https://ui.adsabs.harvard.edu/abs/2013NewAR..57...52B}
}

@ARTICLE{munshi2021,
       author = {{Munshi}, Ferah and {Brooks}, Alyson M. and {Applebaum}, Elaad and {Christensen}, Charlotte R. and {Quinn}, T. and {Sligh}, Serena},
        title = "{Quantifying Scatter in Galaxy Formation at the Lowest Masses}",
      journal = {\apj},
         year = 2021,
        month = dec,
       volume = {923},
       number = {1},
          eid = {35},
        pages = {35},
          doi = {10.3847/1538-4357/ac0db6},
archivePrefix = {arXiv},
       eprint = {2101.05822},
 primaryClass = {astro-ph.GA},
       adsurl = {https://ui.adsabs.harvard.edu/abs/2021ApJ...923...35M}
}

@ARTICLE{nadler2025,
       author = {{Nadler}, Ethan O.},
        title = "{The Impact of Molecular Hydrogen Cooling on the Galaxy Formation Threshold}",
      journal = {\apjl},
         year = 2025,
        month = apr,
       volume = {983},
       number = {1},
          eid = {L23},
        pages = {L23},
          doi = {10.3847/2041-8213/adbc6e},
archivePrefix = {arXiv},
       eprint = {2503.04885},
 primaryClass = {astro-ph.GA},
       adsurl = {https://ui.adsabs.harvard.edu/abs/2025ApJ...983L..23N}
}

@ARTICLE{wheeler2025b,
       author = {{Wheeler}, Vance and {Kravtsov}, Andrey and {Chiti}, Anirudh and {Katz}, Harley and {Semenov}, Vadim A.},
        title = "{What Sets the Metallicity of Ultra-Faint Dwarfs?}",
      journal = {The Open Journal of Astrophysics},
         year = 2025,
        month = oct,
       volume = {8},
          eid = {151},
        pages = {151},
          doi = {10.33232/001c.145734},
archivePrefix = {arXiv},
       eprint = {2507.03182},
 primaryClass = {astro-ph.GA},
       adsurl = {https://ui.adsabs.harvard.edu/abs/2025OJAp....8E.151W}
}

@ARTICLE{euclid2025,
       author = {{Euclid Collaboration} and {Mellier}, Y. and {Abdurro'uf} and {Acevedo Barroso}, J.~A. and {Ach{\'u}carro}, A. and {Adamek}, J. and {Adam}, R. and {Addison}, G.~E. and {Aghanim}, N. and {Aguena}, M. and et al.},
        title = "{Euclid: I. Overview of the Euclid mission}",
      journal = {\aap},
         year = 2025,
        month = may,
       volume = {697},
          eid = {A1},
        pages = {A1},
          doi = {10.1051/0004-6361/202450810},
archivePrefix = {arXiv},
       eprint = {2405.13491},
 primaryClass = {astro-ph.CO},
       adsurl = {https://ui.adsabs.harvard.edu/abs/2025A&A...697A...1E}
}

@ARTICLE{roman2015,
       author = {{Spergel}, D. and {Gehrels}, N. and {Baltay}, C. and {Bennett}, D. and {Breckinridge}, J. and {Donahue}, M. and {Dressler}, A. and {Gaudi}, B.~S. and {Greene}, T. and {Guyon}, O. and et al.},
        title = "{Wide-Field InfrarRed Survey Telescope-Astrophysics Focused Telescope Assets WFIRST-AFTA 2015 Report}",
      journal = {arXiv e-prints},
         year = 2015,
        month = mar,
          eid = {arXiv:1503.03757},
        pages = {arXiv:1503.03757},
          doi = {10.48550/arXiv.1503.03757},
archivePrefix = {arXiv},
       eprint = {1503.03757},
 primaryClass = {astro-ph.IM},
       adsurl = {https://ui.adsabs.harvard.edu/abs/2015arXiv150303757S}
}

@ARTICLE{lsst2019,
       author = {{Ivezi{\'c}}, {\v{Z}}eljko and {Kahn}, Steven M. and {Tyson}, J. Anthony and {Abel}, Bob and {Acosta}, Emily and {Allsman}, Robyn and {Alonso}, David and {AlSayyad}, Yusra and {Anderson}, Scott F. and {Andrew}, John and et al.},
        title = "{LSST: From Science Drivers to Reference Design and Anticipated Data Products}",
      journal = {\apj},
         year = 2019,
        month = mar,
       volume = {873},
       number = {2},
          eid = {111},
        pages = {111},
          doi = {10.3847/1538-4357/ab042c},
archivePrefix = {arXiv},
       eprint = {0805.2366},
 primaryClass = {astro-ph},
       adsurl = {https://ui.adsabs.harvard.edu/abs/2019ApJ...873..111I}
}

@ARTICLE{Abdo2010,
       author = {{Abdo}, A.~A. and {Ackermann}, M. and {Ajello}, M. and {Atwood}, W.~B. and {Baldini}, L. and {Ballet}, J. and {Barbiellini}, G. and {Bastieri}, D. and {Bechtol}, K. and {Bellazzini}, R. and et al.},
        title = "{Observations of Milky Way Dwarf Spheroidal Galaxies with the Fermi-Large Area Telescope Detector and Constraints on Dark Matter Models}",
      journal = {\apj},
         year = 2010,
        month = mar,
       volume = {712},
       number = {1},
        pages = {147-158},
          doi = {10.1088/0004-637X/712/1/147},
archivePrefix = {arXiv},
       eprint = {1001.4531},
 primaryClass = {astro-ph.CO},
       adsurl = {https://ui.adsabs.harvard.edu/abs/2010ApJ...712..147A}
}

@ARTICLE{Strigari2018,
       author = {{Strigari}, Louis E.},
        title = "{Dark matter in dwarf spheroidal galaxies and indirect detection: a review}",
      journal = {Reports on Progress in Physics},
         year = 2018,
        month = may,
       volume = {81},
       number = {5},
          eid = {056901},
        pages = {056901},
          doi = {10.1088/1361-6633/aaae16},
archivePrefix = {arXiv},
       eprint = {1805.05883},
 primaryClass = {astro-ph.CO},
       adsurl = {https://ui.adsabs.harvard.edu/abs/2018RPPh...81e6901S}
}

@ARTICLE{4MOST2023,
       author = {{Sk{\'u}lad{\'o}ttir}, {\'A}. and {Puls}, A.~A. and {Amarsi}, A.~M. and {Battaglia}, G. and {Buder}, S. and {Campbell}, S. and {Cardona-Barrero}, S. and {Christlieb}, N. and {Feuillet}, D.~K. and {Gelli}, V. and {Hansen}, C.~J. and {Hill}, V. and {Ibata}, R. and {Jablonka}, P. and {Kacharov}, N. and {Karakas}, A. and {Koch-Hansen}, A.~J. and {Lind}, K. and {Lombardo}, L. and {Lucchesi}, R.~E.~R. and {Lugaro}, M. and {Martin}, N. and {Massari}, D. and {Nordlander}, T. and {Reichert}, M. and {Rossi}, M. and {Ruiter}, A.~J. and {Salvadori}, S. and {Seitenzahl}, I.~R. and {Tolstoy}, E. and {Xylakis-Dornbusch}, T. and {Youakim}, K.~C.},
        title = "{The 4MOST Survey of Dwarf Galaxies and their Stellar Streams (4DWARFS)}",
      journal = {The Messenger},
         year = 2023,
        month = mar,
       volume = {190},
        pages = {19-21},
          doi = {10.18727/0722-6691/5304},
       adsurl = {https://ui.adsabs.harvard.edu/abs/2023Msngr.190...19S}
}

@ARTICLE{Navabi2025,
       author = {{Navabi}, M. and {Carrera}, R. and {No{\"e}l}, N.~E.~D. and {Gallart}, C. and {Pancino}, E. and {De Leo}, M.},
        title = "{Revisiting the near-infrared calcium triplet as metallicity indicator}",
      journal = {\mnras},
         year = 2026,
        month = feb,
       volume = {546},
       number = {2},
          eid = {stag019},
        pages = {stag019},
          doi = {10.1093/mnras/stag019},
       adsurl = {https://ui.adsabs.harvard.edu/abs/2026MNRAS.546ag019N}
}

@ARTICLE{simon2024,
       author = {{Simon}, Joshua D. and {Li}, Ting S. and {Ji}, Alexander P. and {Pace}, Andrew B. and {Hansen}, Terese T. and {Cerny}, William and {Escala}, Ivanna and {Koposov}, Sergey E. and {Drlica-Wagner}, Alex and {Mau}, Sidney and et al.},
        title = "{Eridanus III and DELVE 1: Carbon-rich Primordial Star Clusters or the Smallest Dwarf Galaxies?}",
      journal = {\apj},
         year = 2024,
        month = dec,
       volume = {976},
       number = {2},
          eid = {256},
        pages = {256},
          doi = {10.3847/1538-4357/ad85dd},
archivePrefix = {arXiv},
       eprint = {2410.08276},
 primaryClass = {astro-ph.GA},
       adsurl = {https://ui.adsabs.harvard.edu/abs/2024ApJ...976..256S}
}

@software{bovy2014,
       author = {{Bovy}, Jo},
        title = "{galpy: Galactic dynamics package}",
 howpublished = {Astrophysics Source Code Library, record ascl:1411.008},
         year = 2014,
        month = nov,
          eid = {ascl:1411.008},
archivePrefix = {ascl},
       eprint = {1411.008},
       adsurl = {https://ui.adsabs.harvard.edu/abs/2014ascl.soft11008B}
}

@ARTICLE{bastian2018,
       author = {{Bastian}, Nate and {Lardo}, Carmela},
        title = "{Multiple Stellar Populations in Globular Clusters}",
      journal = {\araa},
         year = 2018,
        month = sep,
       volume = {56},
        pages = {83-136},
          doi = {10.1146/annurev-astro-081817-051839},
archivePrefix = {arXiv},
       eprint = {1712.01286},
 primaryClass = {astro-ph.SR},
       adsurl = {https://ui.adsabs.harvard.edu/abs/2018ARA&A..56...83B}
}

@ARTICLE{Latour2025,
       author = {{Latour}, M. and {Kamann}, S. and {Martocchia}, S. and {Husser}, T. -O. and {Saracino}, S. and {Dreizler}, S.},
        title = "{A stellar census in globular clusters with MUSE: Metallicity spread and dispersion among first-population stars}",
      journal = {\aap},
         year = 2025,
        month = feb,
       volume = {694},
          eid = {A248},
        pages = {A248},
          doi = {10.1051/0004-6361/202452420},
archivePrefix = {arXiv},
       eprint = {2501.09558},
 primaryClass = {astro-ph.GA},
       adsurl = {https://ui.adsabs.harvard.edu/abs/2025A&A...694A.248L}
}

@ARTICLE{ou2024,
       author = {{Ou}, Xiaowei and {Chiti}, Anirudh and {Shipp}, Nora and {Simon}, Joshua D. and {Geha}, Marla and {Frebel}, Anna and {Mardini}, Mohammad K. and {Erkal}, Denis and {Necib}, Lina},
        title = "{Signatures of Tidal Disruption of the Hercules Ultrafaint Dwarf Galaxy}",
      journal = {\apj},
         year = 2024,
        month = may,
       volume = {966},
       number = {1},
          eid = {33},
        pages = {33},
          doi = {10.3847/1538-4357/ad2f27},
archivePrefix = {arXiv},
       eprint = {2403.00921},
 primaryClass = {astro-ph.GA},
       adsurl = {https://ui.adsabs.harvard.edu/abs/2024ApJ...966...33O}
}

@ARTICLE{li2019,
       author = {{Li}, T.~S. and {Koposov}, S.~E. and {Zucker}, D.~B. and {Lewis}, G.~F. and {Kuehn}, K. and {Simpson}, J.~D. and {Ji}, A.~P. and {Shipp}, N. and {Mao}, Y. -Y. and {Geha}, M. and {Pace}, A.~B. and {Mackey}, A.~D. and {Allam}, S. and {Tucker}, D.~L. and {Da Costa}, G.~S. and {Erkal}, D. and {Simon}, J.~D. and {Mould}, J.~R. and {Martell}, S.~L. and {Wan}, Z. and {De Silva}, G.~M. and {Bechtol}, K. and {Balbinot}, E. and {Belokurov}, V. and {Bland-Hawthorn}, J. and {Casey}, A.~R. and {Cullinane}, L. and {Drlica-Wagner}, A. and {Sharma}, S. and {Vivas}, A.~K. and {Wechsler}, R.~H. and {Yanny}, B. and {S5 Collaboration}},
        title = "{The southern stellar stream spectroscopic survey (S$^{5}$): Overview, target selection, data reduction, validation, and early science}",
      journal = {\mnras},
         year = 2019,
        month = dec,
       volume = {490},
       number = {3},
        pages = {3508-3531},
          doi = {10.1093/mnras/stz2731},
archivePrefix = {arXiv},
       eprint = {1907.09481},
 primaryClass = {astro-ph.GA},
       adsurl = {https://ui.adsabs.harvard.edu/abs/2019MNRAS.490.3508L}
}

@ARTICLE{walker2023,
       author = {{Walker}, Matthew G. and {Caldwell}, Nelson and {Mateo}, Mario and {Olszewski}, Edward W. and {Pace}, Andrew B. and {Bailey}, John I. and {Koposov}, Sergey E. and {Roederer}, Ian U.},
        title = "{Magellan/M2FS and MMT/Hectochelle Spectroscopy of Dwarf Galaxies and Faint Star Clusters within the Galactic Halo}",
      journal = {\apjs},
         year = 2023,
        month = sep,
       volume = {268},
       number = {1},
          eid = {19},
        pages = {19},
          doi = {10.3847/1538-4365/acdd79},
archivePrefix = {arXiv},
       eprint = {2312.12738},
 primaryClass = {astro-ph.GA},
       adsurl = {https://ui.adsabs.harvard.edu/abs/2023ApJS..268...19W}
}

@ARTICLE{Cooper2023,
       author = {{Cooper}, Andrew P. and {Koposov}, Sergey E. and {Allende Prieto}, Carlos and {Manser}, Christopher J. and {Kizhuprakkat}, Namitha and {Myers}, Adam D. and {Dey}, Arjun and {G{\"a}nsicke}, Boris T. and {Li}, Ting S. and {Rockosi}, Constance and {Valluri}, Monica and {Najita}, Joan and {Deason}, Alis and {Raichoor}, Anand and {Wang}, M. -Y. and {Ting}, Y. -S. and {Kim}, Bokyoung and {Carrillo}, Andreia and {Wang}, Wenting and {Beraldo e Silva}, Leandro and {Han}, Jiwon Jesse and {Ding}, Jiani and {S{\'a}nchez-Conde}, Miguel and {Aguilar}, Jessica N. and {Ahlen}, Steven and {Bailey}, Stephen and {Belokurov}, Vasily and {Brooks}, David and {Cunha}, Katia and {Dawson}, Kyle and {de la Macorra}, Axel and {Doel}, Peter and {Eisenstein}, Daniel J. and {Fagrelius}, Parker and {Fanning}, Kevin and {Font-Ribera}, Andreu and {Forero-Romero}, Jaime E. and {Gazta{\~n}aga}, Enrique and {Gontcho a Gontcho}, Satya and {Guy}, Julien and {Honscheid}, Klaus and {Kehoe}, Robert and {Kisner}, Theodore and {Kremin}, Anthony and {Landriau}, Martin and {Levi}, Michael E. and {Martini}, Paul and {Meisner}, Aaron M. and {Miquel}, Ramon and {Moustakas}, John and {Nie}, Jundan J.~D. and {Palanque-Delabrouille}, Nathalie and {Percival}, Will J. and {Poppett}, Claire and {Prada}, Francisco and {Rehemtulla}, Nabeel and {Schlafly}, Edward and {Schlegel}, David and {Schubnell}, Michael and {Sharples}, Ray M. and {Tarl{\'e}}, Gregory and {Wechsler}, Risa H. and {Weinberg}, David H. and {Zhou}, Zhimin and {Zou}, Hu},
        title = "{Overview of the DESI Milky Way Survey}",
      journal = {\apj},
         year = 2023,
        month = apr,
       volume = {947},
       number = {1},
          eid = {37},
        pages = {37},
          doi = {10.3847/1538-4357/acb3c0},
archivePrefix = {arXiv},
       eprint = {2208.08514},
 primaryClass = {astro-ph.GA},
       adsurl = {https://ui.adsabs.harvard.edu/abs/2023ApJ...947...37C}
}

@ARTICLE{koposov2025,
       author = {{Koposov}, Sergey E. and {Li}, Ting S. and {Allende Prieto}, C. and {Medina}, G.~E. and {Sandford}, N. and {Aguado}, D. and {Silva}, L. Beraldo e and {Bystr{\"o}m}, A. and {Cooper}, A.~P. and {Dey}, Arjun and {Frenk}, C.~S. and {Kizhuprakkat}, N. and {Li}, S. and {Najita}, J. and {Riley}, A.~H. and {Silva}, D.~R. and {Thomas}, G. and {Valluri}, M. and {Aguilar}, J. and {Ahlen}, S. and {Bianchi}, D. and {Brooks}, D. and {Claybaugh}, T. and {Cole}, S. and {Cuceu}, A. and {de la Macorra}, A. and {Della Costa}, J. and {Dey}, Biprateep and {Doel}, P. and {Edelstein}, J. and {Font-Ribera}, A. and {Forero-Romero}, J.~E. and {Gazta{\~n}aga}, E. and {Gontcho}, S. Gontcho A and {Gutierrez}, G. and {Guy}, J. and {Honscheid}, K. and {Jimenez}, J. and {Kehoe}, R. and {Kirkby}, D. and {Kisner}, T. and {Kremin}, A. and {Lahav}, O. and {Landriau}, M. and {Le Guillou}, L. and {Leauthaud}, A. and {Levi}, M.~E. and {Manera}, M. and {Meisner}, A. and {Miquel}, R. and {Moustakas}, J. and {Nadathur}, S. and {Palanque-Delabrouille}, N. and {Percival}, W.~J. and {Prada}, F. and {P{\'e}rez-R{\`a}fols}, I. and {Rossi}, G. and {Sanchez}, E. and {Schlafly}, E.~F. and {Schlegel}, D. and {Seo}, H. and {Sharples}, R. and {Silber}, J. and {Sprayberry}, D. and {Tarl'e}, G. and {Weaver}, B.~A. and {Zhou}, R. and {Zou}, H.},
        title = "{DESI Data Release 1: Stellar Catalogue}",
      journal = {arXiv e-prints},
         year = 2025,
        month = may,
          eid = {arXiv:2505.14787},
        pages = {arXiv:2505.14787},
archivePrefix = {arXiv},
       eprint = {2505.14787},
 primaryClass = {astro-ph.GA},
       adsurl = {https://ui.adsabs.harvard.edu/abs/2025arXiv250514787K}
}

@ARTICLE{cerny2024,
       author = {{Cerny}, W. and {Chiti}, A. and {Geha}, M. and {Mutlu-Pakdil}, B. and {Drlica-Wagner}, A. and {Tan}, C.~Y. and {Adam{\'o}w}, M. and {Pace}, A.~B. and {Simon}, J.~D. and {Sand}, D.~J. and {Ji}, A.~P. and {Li}, T.~S. and {Vivas}, A.~K. and {Bell}, E.~F. and {Carlin}, J.~L. and {Carballo-Bello}, J.~A. and {Chaturvedi}, A. and {Choi}, Y. and {Doliva-Dolinsky}, A. and {Gnedin}, O.~Y. and {Limberg}, G. and {Mart{\'\i}nez-V{\'a}zquez}, C.~E. and {Mau}, S. and {Medina}, G.~E. and {Navabi}, M. and {No{\"e}l}, N.~E.~D. and {Placco}, V.~M. and {Riley}, A.~H. and {Roederer}, I.~U. and {Stringfellow}, G.~S. and {Bom}, C.~R. and {Ferguson}, P.~S. and {James}, D.~J. and {Mart{\'\i}nez-Delgado}, D. and {Massana}, P. and {Nidever}, D.~L. and {Sakowska}, J.~D. and {Santana-Silva}, L. and {Sherman}, N.~F. and {Tollerud}, E.~J.},
        title = "{Discovery and Spectroscopic Confirmation of Aquarius III: A Low-mass Milky Way Satellite Galaxy}",
      journal = {\apj},
         year = 2025,
        month = feb,
       volume = {979},
       number = {2},
          eid = {164},
        pages = {164},
          doi = {10.3847/1538-4357/ad8eba},
archivePrefix = {arXiv},
       eprint = {2410.00981},
 primaryClass = {astro-ph.GA},
       adsurl = {https://ui.adsabs.harvard.edu/abs/2025ApJ...979..164C}
}

@ARTICLE{smith2024,
       author = {{Smith}, Simon E.~T. and {Cerny}, William and {Hayes}, Christian R. and {Sestito}, Federico and {Jensen}, Jaclyn and {McConnachie}, Alan W. and {Geha}, Marla and {Navarro}, Julio F. and {Li}, Ting S. and {Cuillandre}, Jean-Charles and {Errani}, Rapha{\"e}l and {Chambers}, Ken and {Gwyn}, Stephen and {Hammer}, Francois and {Hudson}, Michael J. and {Magnier}, Eugene and {Martin}, Nicolas},
        title = "{The Discovery of the Faintest Known Milky Way Satellite Using UNIONS}",
      journal = {\apj},
         year = 2024,
        month = jan,
       volume = {961},
       number = {1},
          eid = {92},
        pages = {92},
          doi = {10.3847/1538-4357/ad0d9f},
archivePrefix = {arXiv},
       eprint = {2311.10147},
 primaryClass = {astro-ph.GA},
       adsurl = {https://ui.adsabs.harvard.edu/abs/2024ApJ...961...92S}
}

@ARTICLE{pace2020,
       author = {{Pace}, Andrew B. and {Kaplinghat}, Manoj and {Kirby}, Evan and {Simon}, Joshua D. and {Tollerud}, Erik and {Mu{\~n}oz}, Ricardo R. and {C{\^o}t{\'e}}, Patrick and {Djorgovski}, S.~G. and {Geha}, Marla},
        title = "{Multiple chemodynamic stellar populations of the Ursa Minor dwarf spheroidal galaxy}",
      journal = {\mnras},
         year = 2020,
        month = jul,
       volume = {495},
       number = {3},
        pages = {3022-3040},
          doi = {10.1093/mnras/staa1419},
archivePrefix = {arXiv},
       eprint = {2002.09503},
 primaryClass = {astro-ph.GA},
       adsurl = {https://ui.adsabs.harvard.edu/abs/2020MNRAS.495.3022P}
}

@ARTICLE{pace2022,
       author = {{Pace}, Andrew B. and {Erkal}, Denis and {Li}, Ting S.},
        title = "{Proper Motions, Orbits, and Tidal Influences of Milky Way Dwarf Spheroidal Galaxies}",
      journal = {\apj},
         year = 2022,
        month = dec,
       volume = {940},
       number = {2},
          eid = {136},
        pages = {136},
          doi = {10.3847/1538-4357/ac997b},
archivePrefix = {arXiv},
       eprint = {2205.05699},
 primaryClass = {astro-ph.GA},
       adsurl = {https://ui.adsabs.harvard.edu/abs/2022ApJ...940..136P}
}

@ARTICLE{pace2024,
 author = {{Pace}, Andrew B},
        title = "{The Local Volume Database: a library of the observed properties of nearby dwarf galaxies and star clusters}",
    journal = {The Open Journal of Astrophysics},
        year = 2025,
        month = sep,
    volume = {8},
        eid = {142},
        pages = {142},
        doi = {10.33232/001c.144859},
    archivePrefix = {arXiv},
    eprint = {2411.07424},
    primaryClass = {astro-ph.GA},
    adsurl = {https://ui.adsabs.harvard.edu/abs/2025OJAp....8E.142P}
}

@ARTICLE{DYNESTY2020,
       author = {{Speagle}, Joshua S.},
        title = "{DYNESTY: a dynamic nested sampling package for estimating Bayesian posteriors and evidences}",
      journal = {\mnras},
         year = 2020,
        month = apr,
       volume = {493},
       number = {3},
        pages = {3132-3158},
          doi = {10.1093/mnras/staa278},
archivePrefix = {arXiv},
       eprint = {1904.02180},
 primaryClass = {astro-ph.IM},
       adsurl = {https://ui.adsabs.harvard.edu/abs/2020MNRAS.493.3132S}
}

@ARTICLE{Walker2006,
       author = {{Walker}, Matthew G. and {Mateo}, Mario and {Olszewski}, Edward W. and {Bernstein}, Rebecca and {Wang}, Xiao and {Woodroofe}, Michael},
        title = "{Internal Kinematics of the Fornax Dwarf Spheroidal Galaxy}",
      journal = {\aj},
         year = 2006,
        month = apr,
       volume = {131},
       number = {4},
        pages = {2114-2139},
          doi = {10.1086/500193},
archivePrefix = {arXiv},
       eprint = {astro-ph/0511465},
 primaryClass = {astro-ph},
       adsurl = {https://ui.adsabs.harvard.edu/abs/2006AJ....131.2114W}
}

@article{Kass95,
author = {{Kass}, Robert E. and  {Raftery}, Adrian E.},
title = {Bayes Factors},
journal = {Journal of the American Statistical Association},
volume = {90},
number = {430},
pages = {773-795},
year  = {1995},
publisher = {Taylor & Francis},
doi = {10.1080/01621459.1995.10476572},
URL = {https://www.tandfonline.com/doi/abs/10.1080/01621459.1995.10476572},
eprint = {https://www.tandfonline.com/doi/pdf/10.1080/01621459.1995.10476572}
}

@ARTICLE{Kravtsov2013,
       author = {{Kravtsov}, Andrey V.},
        title = "{The Size-Virial Radius Relation of Galaxies}",
      journal = {\apjl},
         year = 2013,
        month = feb,
       volume = {764},
       number = {2},
          eid = {L31},
        pages = {L31},
          doi = {10.1088/2041-8205/764/2/L31},
archivePrefix = {arXiv},
       eprint = {1212.2980},
 primaryClass = {astro-ph.CO},
       adsurl = {https://ui.adsabs.harvard.edu/abs/2013ApJ...764L..31K}
}

@ARTICLE{pace2025,
       author = {{Pace}, Andrew B. and {Li}, T.~S. and {Ji}, A.~P. and {Simon}, J.~D. and {Cerny}, W. and {Senkevich}, A.~M. and {Drlica-Wagner}, A. and {Bechtol}, K. and {Tan}, C.~Y. and {Chiti}, A. and et al.},
        title = "{Spectroscopic Analysis of Pictor II: a very low metallicity ultra-faint dwarf galaxy bound to the Large Magellanic Cloud}",
      journal = {The Open Journal of Astrophysics},
         year = 2025,
        month = aug,
       volume = {8},
          eid = {112},
        pages = {112},
          doi = {10.33232/001c.142989},
archivePrefix = {arXiv},
       eprint = {2506.21841},
 primaryClass = {astro-ph.GA},
       adsurl = {https://ui.adsabs.harvard.edu/abs/2025OJAp....8E.112P}
}

@ARTICLE{Zonoozi2024,
       author = {{Zonoozi}, Akram Hasani and {Rabiee}, Maliheh and {Haghi}, Hosein and {Kroupa}, Pavel},
        title = "{Has the Palomar 14 Globular Cluster Been Captured by the Milky Way?}",
      journal = {\apj},
         year = 2024,
        month = nov,
       volume = {975},
       number = {2},
          eid = {266},
        pages = {266},
          doi = {10.3847/1538-4357/ad7953},
archivePrefix = {arXiv},
       eprint = {2410.06036},
 primaryClass = {astro-ph.GA},
       adsurl = {https://ui.adsabs.harvard.edu/abs/2024ApJ...975..266Z}
}

@ARTICLE{Zaremba2025,
       author = {{Zaremba}, Daria and {Venn}, Kim and {Hayes}, Christian R. and {Errani}, Rapha{\"e}l and {Cornejo}, Triana and {Glover}, Jennifer and {Jensen}, Jaclyn and {McConnachie}, Alan W. and {Navarro}, Julio F. and {Pazder}, John and et al.},
        title = "{GHOST Commissioning Science Results. IV. Chemodynamical Analyses of Milky Way Satellites Sagittarius II and Aquarius II}",
      journal = {\apj},
         year = 2025,
        month = jul,
       volume = {987},
       number = {2},
          eid = {217},
        pages = {217},
          doi = {10.3847/1538-4357/add5f9},
archivePrefix = {arXiv},
       eprint = {2503.05927},
 primaryClass = {astro-ph.GA},
       adsurl = {https://ui.adsabs.harvard.edu/abs/2025ApJ...987..217Z}
}

@ARTICLE{bailin2019,
       author = {{Bailin}, Jeremy},
        title = "{Globular Cluster Intrinsic Iron Abundance Spreads. I. Catalog}",
      journal = {\apjs},
         year = 2019,
        month = nov,
       volume = {245},
       number = {1},
          eid = {5},
        pages = {5},
          doi = {10.3847/1538-4365/ab4812},
archivePrefix = {arXiv},
       eprint = {1909.11731},
 primaryClass = {astro-ph.GA},
       adsurl = {https://ui.adsabs.harvard.edu/abs/2019ApJS..245....5B}
}

@ARTICLE{Tsantaki2022,
       author = {{Tsantaki}, M. and {Pancino}, E. and {Marrese}, P. and {Marinoni}, S. and {Rainer}, M. and {Sanna}, N. and {Turchi}, A. and {Randich}, S. and {Gallart}, C. and {Battaglia}, G. and {Masseron}, T.},
        title = "{Survey of Surveys. I. The largest compilation of radial velocities for the Galaxy}",
      journal = {\aap},
         year = 2022,
        month = mar,
       volume = {659},
          eid = {A95},
        pages = {A95},
          doi = {10.1051/0004-6361/202141702},
archivePrefix = {arXiv},
       eprint = {2110.09316},
 primaryClass = {astro-ph.GA},
       adsurl = {https://ui.adsabs.harvard.edu/abs/2022A&A...659A..95T}
}

@ARTICLE{pace2023,
       author = {{Pace}, Andrew B. and {Koposov}, Sergey E. and {Walker}, Matthew G. and {Caldwell}, Nelson and {Mateo}, Mario and {Olszewski}, Edward W. and {Roederer}, Ian U. and {Bailey}, John I. and {Belokurov}, Vasily and {Kuehn}, Kyler and {Li}, Ting S. and {Zucker}, Daniel B.},
        title = "{The kinematics, metallicities, and orbits of six recently discovered Galactic star clusters with Magellan/M2FS spectroscopy}",
      journal = {\mnras},
         year = 2023,
        month = nov,
       volume = {526},
       number = {1},
        pages = {1075-1094},
          doi = {10.1093/mnras/stad2760},
archivePrefix = {arXiv},
       eprint = {2304.06904},
 primaryClass = {astro-ph.GA},
       adsurl = {https://ui.adsabs.harvard.edu/abs/2023MNRAS.526.1075P}
}

@ARTICLE{buttry2022,
       author = {{Buttry}, Rachel and {Pace}, Andrew B. and {Koposov}, Sergey E. and {Walker}, Matthew G. and {Caldwell}, Nelson and {Kirby}, Evan N. and {Martin}, Nicolas F. and {Mateo}, Mario and {Olszewski}, Edward W. and {Starkenburg}, Else and {Badenes}, Carles and {Daher}, Christine Mazzola},
        title = "{Stellar kinematics of dwarf galaxies from multi-epoch spectroscopy: application to Triangulum II}",
      journal = {\mnras},
         year = 2022,
        month = aug,
       volume = {514},
       number = {2},
        pages = {1706-1719},
          doi = {10.1093/mnras/stac1441},
archivePrefix = {arXiv},
       eprint = {2108.10867},
 primaryClass = {astro-ph.GA},
       adsurl = {https://ui.adsabs.harvard.edu/abs/2022MNRAS.514.1706B}
}

@ARTICLE{Baumgardt2019,
       author = {{Baumgardt}, H. and {Hilker}, M. and {Sollima}, A. and {Bellini}, A.},
        title = "{Mean proper motions, space orbits, and velocity dispersion profiles of Galactic globular clusters derived from Gaia DR2 data}",
      journal = {\mnras},
         year = 2019,
        month = feb,
       volume = {482},
       number = {4},
        pages = {5138-5155},
          doi = {10.1093/mnras/sty2997},
archivePrefix = {arXiv},
       eprint = {1811.01507},
 primaryClass = {astro-ph.GA},
       adsurl = {https://ui.adsabs.harvard.edu/abs/2019MNRAS.482.5138B}
}

@ARTICLE{pypeit2020,
       author = {{Prochaska}, J. and {Hennawi}, Joseph and {Westfall}, Kyle and {Cooke}, Ryan and {Wang}, Feige and {Hsyu}, Tiffany and {Davies}, Frederick and {Farina}, Emanuele and {Pelliccia}, Debora},
        title = "{PypeIt: The Python Spectroscopic Data Reduction Pipeline}",
      journal = {The Journal of Open Source Software},
         year = 2020,
        month = dec,
       volume = {5},
       number = {56},
          eid = {2308},
        pages = {2308},
          doi = {10.21105/joss.02308},
archivePrefix = {arXiv},
       eprint = {2005.06505},
 primaryClass = {astro-ph.IM},
       adsurl = {https://ui.adsabs.harvard.edu/abs/2020JOSS....5.2308P}
}

@ARTICLE{emcee,
       author = {{Foreman-Mackey}, Daniel and {Hogg}, David W. and {Lang}, Dustin and
         {Goodman}, Jonathan},
        title = "{emcee: The MCMC Hammer}",
      journal = {\pasp},
         year = 2013,
        month = mar,
       volume = {125},
       number = {925},
        pages = {306},
          doi = {10.1086/670067},
archivePrefix = {arXiv},
       eprint = {1202.3665},
 primaryClass = {astro-ph.IM},
       adsurl = {https://ui.adsabs.harvard.edu/abs/2013PASP..125..306F}
}

@ARTICLE{rey2025,
       author = {{Rey}, Martin P. and {Taylor}, Ethan and {Gray}, Emily I. and {Kim}, Stacy Y. and {Andersson}, Eric P. and {Pontzen}, Andrew and {Agertz}, Oscar and {Read}, Justin I. and {Cadiou}, Corentin and {Yates}, Robert M. and {Orkney}, Matthew D.~A. and {Scholte}, Dirk and {Saintonge}, Am{\'e}lie and {Breneman}, Joseph and {McQuinn}, Kristen B.~W. and {Muni}, Claudia and {Das}, Payel},
        title = "{EDGE: the emergence of dwarf galaxy scaling relations from cosmological radiation-hydrodynamics simulations}",
      journal = {\mnras},
         year = 2025,
        month = aug,
       volume = {541},
       number = {2},
        pages = {1195-1217},
          doi = {10.1093/mnras/staf1058},
archivePrefix = {arXiv},
       eprint = {2503.03813},
 primaryClass = {astro-ph.GA},
       adsurl = {https://ui.adsabs.harvard.edu/abs/2025MNRAS.541.1195R}
}

@ARTICLE{Leitinger2025,
       author = {{Leitinger}, E.~I. and {Baumgardt}, H. and {Cabrera-Ziri}, I. and {Hilker}, M. and {Carbajo-Hijarrubia}, J. and {Gieles}, M. and {Husser}, T.~O. and {Kamann}, S.},
        title = "{The kinematics of 30 Milky Way globular clusters and the multiple stellar populations within}",
      journal = {\aap},
         year = 2025,
        month = feb,
       volume = {694},
          eid = {A184},
        pages = {A184},
          doi = {10.1051/0004-6361/202452477},
archivePrefix = {arXiv},
       eprint = {2410.02855},
 primaryClass = {astro-ph.GA},
       adsurl = {https://ui.adsabs.harvard.edu/abs/2025A&A...694A.184L}
}

@ARTICLE{segue2008,
       author = {{Adelman-McCarthy}, Jennifer K. and {Ag{\"u}eros}, Marcel A. and {Allam}, Sahar S. and {Allende Prieto}, Carlos and {Anderson}, Kurt S.~J. and {Anderson}, Scott F. and {Annis}, James and {Bahcall}, Neta A. and {Bailer-Jones}, C.~A.~L. and {Baldry}, Ivan K. and {Barentine}, J.~C. and {Bassett}, Bruce A. and {Becker}, Andrew C. and {Beers}, Timothy C. and {Bell}, Eric F. and {Berlind}, Andreas A. and {Bernardi}, Mariangela and {Blanton}, Michael R. and {Bochanski}, John J. and {Boroski}, William N. and {Brinchmann}, Jarle and {Brinkmann}, J. and {Brunner}, Robert J. and {Budav{\'a}ri}, Tam{\'a}s and {Carliles}, Samuel and {Carr}, Michael A. and {Castander}, Francisco J. and {Cinabro}, David and {Cool}, R.~J. and {Covey}, Kevin R. and {Csabai}, Istv{\'a}n and {Cunha}, Carlos E. and {Davenport}, James R.~A. and {Dilday}, Ben and {Doi}, Mamoru and {Eisenstein}, Daniel J. and {Evans}, Michael L. and {Fan}, Xiaohui and {Finkbeiner}, Douglas P. and {Friedman}, Scott D. and {Frieman}, Joshua A. and {Fukugita}, Masataka and {G{\"a}nsicke}, Boris T. and {Gates}, Evalyn and {Gillespie}, Bruce and {Glazebrook}, Karl and {Gray}, Jim and {Grebel}, Eva K. and {Gunn}, James E. and {Gurbani}, Vijay K. and {Hall}, Patrick B. and {Harding}, Paul and {Harvanek}, Michael and {Hawley}, Suzanne L. and {Hayes}, Jeffrey and {Heckman}, Timothy M. and {Hendry}, John S. and {Hindsley}, Robert B. and {Hirata}, Christopher M. and {Hogan}, Craig J. and {Hogg}, David W. and {Hyde}, Joseph B. and {Ichikawa}, Shin-ichi and {Ivezi{\'c}}, {\v{Z}}eljko and {Jester}, Sebastian and {Johnson}, Jennifer A. and {Jorgensen}, Anders M. and {Juri{\'c}}, Mario and {Kent}, Stephen M. and {Kessler}, R. and {Kleinman}, S.~J. and {Knapp}, G.~R. and {Kron}, Richard G. and {Krzesinski}, Jurek and {Kuropatkin}, Nikolay and {Lamb}, Donald Q. and {Lampeitl}, Hubert and {Lebedeva}, Svetlana and {Lee}, Young Sun and {French Leger}, R. and {L{\'e}pine}, S{\'e}bastien and {Lima}, Marcos and {Lin}, Huan and {Long}, Daniel C. and {Loomis}, Craig P. and {Loveday}, Jon and {Lupton}, Robert H. and {Malanushenko}, Olena and {Malanushenko}, Viktor and {Mandelbaum}, Rachel and {Margon}, Bruce and {Marriner}, John P. and {Mart{\'\i}nez-Delgado}, David and {Matsubara}, Takahiko and {McGehee}, Peregrine M. and {McKay}, Timothy A. and {Meiksin}, Avery and {Morrison}, Heather L. and {Munn}, Jeffrey A. and {Nakajima}, Reiko and {Neilsen}, Jr., Eric H. and {Newberg}, Heidi Jo and {Nichol}, Robert C. and {Nicinski}, Tom and {Nieto-Santisteban}, Maria and {Nitta}, Atsuko and {Okamura}, Sadanori and {Owen}, Russell and {Oyaizu}, Hiroaki and {Padmanabhan}, Nikhil and {Pan}, Kaike and {Park}, Changbom and {Peoples}, Jr., John and {Pier}, Jeffrey R. and {Pope}, Adrian C. and {Purger}, Norbert and {Raddick}, M. Jordan and {Re Fiorentin}, Paola and {Richards}, Gordon T. and {Richmond}, Michael W. and {Riess}, Adam G. and {Rix}, Hans-Walter and {Rockosi}, Constance M. and {Sako}, Masao and {Schlegel}, David J. and {Schneider}, Donald P. and {Schreiber}, Matthias R. and {Schwope}, Axel D. and {Seljak}, Uro{\v{s}} and {Sesar}, Branimir and {Sheldon}, Erin and {Shimasaku}, Kazu and {Sivarani}, Thirupathi and {Allyn Smith}, J. and {Snedden}, Stephanie A. and {Steinmetz}, Matthias and {Strauss}, Michael A. and {SubbaRao}, Mark and {Suto}, Yasushi and {Szalay}, Alexander S. and {Szapudi}, Istv{\'a}n and {Szkody}, Paula and {Tegmark}, Max and {Thakar}, Aniruddha R. and {Tremonti}, Christy A. and {Tucker}, Douglas L. and {Uomoto}, Alan and {Vanden Berk}, Daniel E. and {Vandenberg}, Jan and {Vidrih}, S. and {Vogeley}, Michael S. and {Voges}, Wolfgang and {Vogt}, Nicole P. and {Wadadekar}, Yogesh and {Weinberg}, David H. and {West}, Andrew A. and {White}, Simon D.~M. and {Wilhite}, Brian C. and {Yanny}, Brian and {Yocum}, D.~R. and {York}, Donald G. and {Zehavi}, Idit and {Zucker}, Daniel B.},
        title = "{The Sixth Data Release of the Sloan Digital Sky Survey}",
      journal = {\apjs},
         year = 2008,
        month = apr,
       volume = {175},
       number = {2},
        pages = {297-313},
          doi = {10.1086/524984},
archivePrefix = {arXiv},
       eprint = {0707.3413},
 primaryClass = {astro-ph},
       adsurl = {https://ui.adsabs.harvard.edu/abs/2008ApJS..175..297A}
}

@ARTICLE{lamost2019,
       author = {{Xiang}, Maosheng and {Ting}, Yuan-Sen and {Rix}, Hans-Walter and {Sandford}, Nathan and {Buder}, Sven and {Lind}, Karin and {Liu}, Xiao-Wei and {Shi}, Jian-Rong and {Zhang}, Hua-Wei},
        title = "{Abundance Estimates for 16 Elements in 6 Million Stars from LAMOST DR5 Low-Resolution Spectra}",
      journal = {\apjs},
         year = 2019,
        month = dec,
       volume = {245},
       number = {2},
          eid = {34},
        pages = {34},
          doi = {10.3847/1538-4365/ab5364},
archivePrefix = {arXiv},
       eprint = {1908.09727},
 primaryClass = {astro-ph.SR},
       adsurl = {https://ui.adsabs.harvard.edu/abs/2019ApJS..245...34X}
}

@ARTICLE{Majewski2017,
   author = {{Majewski}, S.~R. and {Schiavon}, R.~P. and {Frinchaboy}, P.~M. and 
	{Allende Prieto}, C. and {Barkhouser}, R. and {Bizyaev}, D. and 
	{Blank}, B. and {Brunner}, S. and {Burton}, A. and {Carrera}, R. and 
	{Chojnowski}, S.~D. and {Cunha}, K. and {Epstein}, C. and {Fitzgerald}, G. and 
	{Garc{\'{\i}}a P{\'e}rez}, A.~E. and {Hearty}, F.~R. and {Henderson}, C. and 
	{Holtzman}, J.~A. and {Johnson}, J.~A. and {Lam}, C.~R. and 
	{Lawler}, J.~E. and {Maseman}, P. and {M{\'e}sz{\'a}ros}, S. and 
	{Nelson}, M. and {Nguyen}, D.~C. and {Nidever}, D.~L. and {Pinsonneault}, M. and 
	{Shetrone}, M. and {Smee}, S. and {Smith}, V.~V. and {Stolberg}, T. and 
	{Skrutskie}, M.~F. and {Walker}, E. and {Wilson}, J.~C. and 
	{Zasowski}, G. and {Anders}, F. and {Basu}, S. and {Beland}, S. and 
	{Blanton}, M.~R. and {Bovy}, J. and {Brownstein}, J.~R. and 
	{Carlberg}, J. and {Chaplin}, W. and {Chiappini}, C. and {Eisenstein}, D.~J. and 
	{Elsworth}, Y. and {Feuillet}, D. and {Fleming}, S.~W. and {Galbraith-Frew}, J. and 
	{Garc{\'{\i}}a}, R.~A. and {Garc{\'{\i}}a-Hern{\'a}ndez}, D.~A. and 
	{Gillespie}, B.~A. and {Girardi}, L. and {Gunn}, J.~E. and {Hasselquist}, S. and 
	{Hayden}, M.~R. and {Hekker}, S. and {Ivans}, I. and {Kinemuchi}, K. and 
	{Klaene}, M. and {Mahadevan}, S. and {Mathur}, S. and {Mosser}, B. and 
	{Muna}, D. and {Munn}, J.~A. and {Nichol}, R.~C. and {O'Connell}, R.~W. and 
	{Parejko}, J.~K. and {Robin}, A.~C. and {Rocha-Pinto}, H. and 
	{Schultheis}, M. and {Serenelli}, A.~M. and {Shane}, N. and 
	{Silva Aguirre}, V. and {Sobeck}, J.~S. and {Thompson}, B. and 
	{Troup}, N.~W. and {Weinberg}, D.~H. and {Zamora}, O.},
    title = "{The Apache Point Observatory Galactic Evolution Experiment (APOGEE)}",
  journal = {\aj},
archivePrefix = "arXiv",
   eprint = {1509.05420},
 primaryClass = "astro-ph.IM",
     year = 2017,
    month = sep,
   volume = 154,
      eid = {94},
    pages = {94},
      doi = {10.3847/1538-3881/aa784d},
   adsurl = {http://adsabs.harvard.edu/abs/2017AJ....154...94M}
}

@ARTICLE{Harris2010,
   author = {{Harris}, W.~E.},
    title = "{A New Catalog of Globular Clusters in the Milky Way}",
  journal = {arXiv e-prints},
archivePrefix = "arXiv",
   eprint = {1012.3224},
 primaryClass = "astro-ph.GA",
     year = 2010,
    month = dec,
   adsurl = {http://adsabs.harvard.edu/abs/2010arXiv1012.3224H}
}

@ARTICLE{Carrera2013,
   author = {{Carrera}, R. and {Pancino}, E. and {Gallart}, C. and {del Pino}, A.
	},
    title = "{The near-infrared Ca II triplet as a metallicity indicator - II. Extension to extremely metal-poor metallicity regimes}",
  journal = {\mnras},
archivePrefix = "arXiv",
   eprint = {1306.3883},
     year = 2013,
    month = sep,
   volume = 434,
    pages = {1681-1691},
      doi = {10.1093/mnras/stt1126},
   adsurl = {http://adsabs.harvard.edu/abs/2013MNRAS.434.1681C}
}

@ARTICLE{GaiaDR3,
       author = {{Gaia Collaboration} and {Vallenari}, A. and {Brown}, A.~G.~A. and {Prusti}, T. and {de Bruijne}, J.~H.~J. and {others}},
        title = "{Gaia Data Release 3. Summary of the content and survey properties}",
      journal = {\aap},
         year = 2023,
        month = jun,
       volume = {674},
          eid = {A1},
        pages = {A1},
          doi = {10.1051/0004-6361/202243940},
archivePrefix = {arXiv},
       eprint = {2208.00211},
 primaryClass = {astro-ph.GA},
       adsurl = {https://ui.adsabs.harvard.edu/abs/2023A&A...674A...1G}
}

@ARTICLE{Helmi2006,
   author = {{Helmi}, A. and {Irwin}, M.~J. and {Tolstoy}, E. and {Battaglia}, G. and 
	{Hill}, V. and {Jablonka}, P. and {Venn}, K. and {Shetrone}, M. and 
	{Letarte}, B. and {Arimoto}, N. and {Abel}, T. and {Francois}, P. and 
	{Kaufer}, A. and {Primas}, F. and {Sadakane}, K. and {Szeifert}, T.
	},
    title = "{A New View of the Dwarf Spheroidal Satellites of the Milky Way from VLT FLAMES: Where Are the Very Metal-poor Stars?}",
  journal = {\apjl},
   eprint = {astro-ph/0611420},
     year = 2006,
    month = nov,
   volume = 651,
    pages = {L121-L124},
      doi = {10.1086/509784},
   adsurl = {http://adsabs.harvard.edu/abs/2006ApJ...651L.121H}
}

@ARTICLE{Walker2007,
   author = {{Walker}, M.~G. and {Mateo}, M. and {Olszewski}, E.~W. and {Gnedin}, O.~Y. and 
	{Wang}, X. and {Sen}, B. and {Woodroofe}, M.},
    title = "{Velocity Dispersion Profiles of Seven Dwarf Spheroidal Galaxies}",
  journal = {\apjl},
archivePrefix = "arXiv",
   eprint = {0708.0010},
     year = 2007,
    month = sep,
   volume = 667,
    pages = {L53-L56},
      doi = {10.1086/521998},
   adsurl = {http://adsabs.harvard.edu/abs/2007ApJ...667L..53W}
}

@ARTICLE{Kirby2011,
   author = {{Kirby}, E.~N. and {Cohen}, J.~G. and {Smith}, G.~H. and {Majewski}, S.~R. and 
	{Sohn}, S.~T. and {Guhathakurta}, P.},
    title = "{Multi-element Abundance Measurements from Medium-resolution Spectra. IV. Alpha Element Distributions in Milky Way Satellite Galaxies}",
  journal = {\apj},
archivePrefix = "arXiv",
   eprint = {1011.5221},
 primaryClass = "astro-ph.GA",
     year = 2011,
    month = feb,
   volume = 727,
      eid = {79},
    pages = {79},
      doi = {10.1088/0004-637X/727/2/79},
   adsurl = {http://adsabs.harvard.edu/abs/2011ApJ...727...79K}
}

@ARTICLE{Carlin2018,
   author = {{Carlin}, J.~L. and {Sand}, D.~J.},
    title = "{Bo{\"o}tes III is a Disrupting Dwarf Galaxy Associated with the Styx Stellar Stream}",
  journal = {\apj},
archivePrefix = "arXiv",
   eprint = {1805.11624},
     year = 2018,
    month = sep,
   volume = 865,
      eid = {7},
    pages = {7},
      doi = {10.3847/1538-4357/aad8c1},
   adsurl = {http://adsabs.harvard.edu/abs/2018ApJ...865....7C}
}

@INPROCEEDINGS{faber03a,
   author = {{Faber}, S.~M. and others},
    title = "{The DEIMOS spectrograph for the Keck II Telescope: integration and testing}",
booktitle = {Instrument Design and Performance for Optical/Infrared Ground-based Telescopes.  Edited by Iye \& Moorwood, Proceedings of the SPIE, Volume 4841, pp. 1657},
     year = 2003,
    month = mar,
    pages = {},
   adsurl = {http://adsabs.harvard.edu/cgi-bin/nph-bib_query?bibcode=2003SPIE.4841.1657F&db_key=AST}
}

@ARTICLE{kirby10a,
   author = {{Kirby}, E.~N. and {Guhathakurta}, P. and {Simon}, J.~D. and 
	{Geha}, M.~C. and {Rockosi}, C.~M. and {Sneden}, C. and {Cohen}, J.~G. and 
	{Sohn}, S.~T. and {Majewski}, S.~R. and {Siegel}, M.},
    title = "{Multi-element Abundance Measurements from Medium-resolution Spectra. II. Catalog of Stars in Milky Way Dwarf Satellite Galaxies}",
  journal = {\apjs},
archivePrefix = "arXiv",
   eprint = {1011.4516},
 primaryClass = "astro-ph.GA",
     year = 2010,
    month = dec,
   volume = 191,
    pages = {352-375},
      doi = {10.1088/0067-0049/191/2/352},
   adsurl = {http://adsabs.harvard.edu/abs/2010ApJS..191..352K}
}

@ARTICLE{Martin2007,
   author = {{Martin}, N.~F. and {Ibata}, R.~A. and {Chapman}, S.~C. and 
	{Irwin}, M. and {Lewis}, G.~F.},
    title = "{A Keck/DEIMOS spectroscopic survey of faint Galactic satellites: searching for the least massive dwarf galaxies}",
  journal = {\mnras},
archivePrefix = "arXiv",
   eprint = {0705.4622},
     year = 2007,
    month = sep,
   volume = 380,
    pages = {281-300},
      doi = {10.1111/j.1365-2966.2007.12055.x},
   adsurl = {http://adsabs.harvard.edu/abs/2007MNRAS.380..281M}
}

@ARTICLE{simon07a,
   author = {{Simon}, J.~D. and {Geha}, M.},
    title = "{The Kinematics of the Ultra-faint Milky Way Satellites: Solving the Missing Satellite Problem}",
  journal = {\apj},
   eprint = {arXiv:0706.0516},
     year = 2007,
    month = nov,
   volume = 670,
    pages = {313-331},
      doi = {10.1086/521816},
   adsurl = {http://adsabs.harvard.edu/abs/2007ApJ...670..313S}
}

@ARTICLE{wetzel2016,
   author = {{Wetzel}, A.~R. and {Hopkins}, P.~F. and {Kim}, J.-h. and {Faucher-Gigu{\`e}re}, C.-A. and 
	{Kere{\v s}}, D. and {Quataert}, E.},
    title = "{Reconciling Dwarf Galaxies with {$\Lambda$}CDM Cosmology: Simulating a Realistic Population of Satellites around a Milky Way-mass Galaxy}",
  journal = {\apjl},
archivePrefix = "arXiv",
   eprint = {1602.05957},
     year = 2016,
    month = aug,
   volume = 827,
      eid = {L23},
    pages = {L23},
      doi = {10.3847/2041-8205/827/2/L23},
   adsurl = {http://adsabs.harvard.edu/abs/2016ApJ...827L..23W}
}

@ARTICLE{Cohen2010,
       author = {{Cohen}, Judith G. and {Kirby}, Evan N. and {Simon}, Joshua D. and {Geha}, Marla},
        title = "{NGC 2419-Another Remnant of Accretion by the Milky Way}",
      journal = {\apj},
         year = 2010,
        month = dec,
       volume = {725},
       number = {1},
        pages = {288-295},
          doi = {10.1088/0004-637X/725/1/288},
archivePrefix = {arXiv},
       eprint = {1010.0031},
 primaryClass = {astro-ph.GA},
       adsurl = {https://ui.adsabs.harvard.edu/abs/2010ApJ...725..288C}
}

@ARTICLE{larsen2019,
       author = {{Larsen}, S{\o}ren S. and {Baumgardt}, Holger and {Bastian}, Nate and {Hernandez}, Svea and {Brodie}, Jean},
        title = "{Hubble Space Telescope photometry of multiple stellar populations in the inner parts of NGC 2419}",
      journal = {\aap},
         year = 2019,
        month = apr,
       volume = {624},
          eid = {A25},
        pages = {A25},
          doi = {10.1051/0004-6361/201834494},
archivePrefix = {arXiv},
       eprint = {1902.01416},
 primaryClass = {astro-ph.SR},
       adsurl = {https://ui.adsabs.harvard.edu/abs/2019A&A...624A..25L}
}

@ARTICLE{Carretta2009,
       author = {{Carretta}, E. and {Bragaglia}, A. and {Gratton}, R. and {D'Orazi}, V. and {Lucatello}, S.},
        title = "{Intrinsic iron spread and a new metallicity scale for globular clusters}",
      journal = {\aap},
         year = 2009,
        month = dec,
       volume = {508},
       number = {2},
        pages = {695-706},
          doi = {10.1051/0004-6361/200913003},
archivePrefix = {arXiv},
       eprint = {0910.0675},
 primaryClass = {astro-ph.GA},
       adsurl = {https://ui.adsabs.harvard.edu/abs/2009A&A...508..695C}
}

@ARTICLE{Tolstoy2023,
       author = {{Tolstoy}, Eline and {Sk{\'u}lad{\'o}ttir}, {\'A}sa and {Battaglia}, Giuseppina and {Brown}, Anthony G.~A. and {Massari}, Davide and {Irwin}, Michael J. and {Starkenburg}, Else and {Salvadori}, Stefania and {Hill}, Vanessa and {Jablonka}, Pascale and {Salaris}, Maurizio and {van Essen}, Thom and {Olsthoorn}, Carla and {Helmi}, Amina and {Pritchard}, John},
        title = "{A 3D view of dwarf galaxies with Gaia and VLT/FLAMES. I. The Sculptor dwarf spheroidal}",
      journal = {\aap},
         year = 2023,
        month = jul,
       volume = {675},
          eid = {A49},
        pages = {A49},
          doi = {10.1051/0004-6361/202245717},
archivePrefix = {arXiv},
       eprint = {2304.11980},
 primaryClass = {astro-ph.GA},
       adsurl = {https://ui.adsabs.harvard.edu/abs/2023A&A...675A..49T}
}

@ARTICLE{guerra2023,
       author = {{Guerra}, Juan and {Geha}, Marla and {Strigari}, Louis E.},
        title = "{Forecasts on the Dark Matter Density Profiles of Dwarf Spheroidal Galaxies with Current and Future Kinematic Observations}",
      journal = {\apj},
         year = 2023,
        month = feb,
       volume = {943},
       number = {2},
          eid = {121},
        pages = {121},
          doi = {10.3847/1538-4357/aca8a5},
archivePrefix = {arXiv},
       eprint = {2112.05166},
 primaryClass = {astro-ph.GA},
       adsurl = {https://ui.adsabs.harvard.edu/abs/2023ApJ...943..121G}
}

@ARTICLE{pianta2022,
       author = {{Pianta}, Camilla and {Capuzzo-Dolcetta}, Roberto and {Carraro}, Giovanni},
        title = "{The Impact of Binaries on the Dynamical Mass Estimate of Dwarf Galaxies}",
      journal = {\apj},
         year = 2022,
        month = nov,
       volume = {939},
       number = {1},
          eid = {3},
        pages = {3},
          doi = {10.3847/1538-4357/ac9303},
archivePrefix = {arXiv},
       eprint = {2209.08296},
 primaryClass = {astro-ph.GA},
       adsurl = {https://ui.adsabs.harvard.edu/abs/2022ApJ...939....3P}
}

@ARTICLE{review2022,
       author = {{Battaglia}, Giuseppina and {Nipoti}, Carlo},
        title = "{Stellar dynamics and dark matter in Local Group dwarf galaxies}",
      journal = {Nature Astronomy},
         year = 2022,
        month = may,
       volume = {6},
        pages = {659-672},
          doi = {10.1038/s41550-022-01638-7},
archivePrefix = {arXiv},
       eprint = {2205.07821},
 primaryClass = {astro-ph.GA},
       adsurl = {https://ui.adsabs.harvard.edu/abs/2022NatAs...6..659B}
}

@ARTICLE{bullock2017,
       author = {{Bullock}, James S. and {Boylan-Kolchin}, Michael},
        title = "{Small-Scale Challenges to the {\ensuremath{\Lambda}}CDM Paradigm}",
      journal = {\araa},
         year = 2017,
        month = aug,
       volume = {55},
       number = {1},
        pages = {343-387},
          doi = {10.1146/annurev-astro-091916-055313},
archivePrefix = {arXiv},
       eprint = {1707.04256},
 primaryClass = {astro-ph.CO},
       adsurl = {https://ui.adsabs.harvard.edu/abs/2017ARA&A..55..343B}
}

@ARTICLE{simon2019,
       author = {{Simon}, Joshua D.},
        title = "{The Faintest Dwarf Galaxies}",
      journal = {\araa},
         year = 2019,
        month = aug,
       volume = {57},
        pages = {375-415},
          doi = {10.1146/annurev-astro-091918-104453},
archivePrefix = {arXiv},
       eprint = {1901.05465},
 primaryClass = {astro-ph.GA},
       adsurl = {https://ui.adsabs.harvard.edu/abs/2019ARA&A..57..375S}
}

@ARTICLE{nadler2024A,
       author = {{Nadler}, Ethan O. and {Gluscevic}, Vera and {Driskell}, Trey and {Wechsler}, Risa H. and {Moustakas}, Leonidas A. and {Benson}, Andrew and {Mao}, Yao-Yuan},
        title = "{Forecasts for Galaxy Formation and Dark Matter Constraints from Dwarf Galaxy Surveys}",
      journal = {\apj},
         year = 2024,
        month = may,
       volume = {967},
       number = {1},
          eid = {61},
        pages = {61},
          doi = {10.3847/1538-4357/ad3bb1},
archivePrefix = {arXiv},
       eprint = {2401.10318},
 primaryClass = {astro-ph.GA},
       adsurl = {https://ui.adsabs.harvard.edu/abs/2024ApJ...967...61N}
}

@ARTICLE{read2021,
       author = {{Read}, J.~I. and {Mamon}, G.~A. and {Vasiliev}, E. and {Watkins}, L.~L. and {Walker}, M.~G. and {Pe{\~n}arrubia}, J. and {Wilkinson}, M. and {Dehnen}, W. and {Das}, P.},
        title = "{Breaking beta: a comparison of mass modelling methods for spherical systems}",
      journal = {\mnras},
         year = 2021,
        month = feb,
       volume = {501},
       number = {1},
        pages = {978-993},
          doi = {10.1093/mnras/staa3663},
archivePrefix = {arXiv},
       eprint = {2011.09493},
 primaryClass = {astro-ph.GA},
       adsurl = {https://ui.adsabs.harvard.edu/abs/2021MNRAS.501..978R}
}

@ARTICLE{Baumgardt2018,
       author = {{Baumgardt}, H. and {Hilker}, M.},
        title = "{A catalogue of masses, structural parameters, and velocity dispersion profiles of 112 Milky Way globular clusters}",
      journal = {\mnras},
         year = 2018,
        month = aug,
       volume = {478},
       number = {2},
        pages = {1520-1557},
          doi = {10.1093/mnras/sty1057},
archivePrefix = {arXiv},
       eprint = {1804.08359},
 primaryClass = {astro-ph.GA},
       adsurl = {https://ui.adsabs.harvard.edu/abs/2018MNRAS.478.1520B}
}

@ARTICLE{sollima2019,
       author = {{Sollima}, A. and {Baumgardt}, H. and {Hilker}, M.},
        title = "{The eye of Gaia on globular clusters kinematics: internal rotation}",
      journal = {\mnras},
         year = 2019,
        month = may,
       volume = {485},
       number = {1},
        pages = {1460-1476},
          doi = {10.1093/mnras/stz505},
archivePrefix = {arXiv},
       eprint = {1902.05895},
 primaryClass = {astro-ph.GA},
       adsurl = {https://ui.adsabs.harvard.edu/abs/2019MNRAS.485.1460S}
}

@ARTICLE{Tan2025,
       author = {{Tan}, C.~Y. and {Cerny}, W. and {Drlica-Wagner}, A. and {Pace}, A.~B. and {Geha}, M. and {Ji}, A.~P. and {Li}, T.~S. and {Adam{\'o}w}, M. and {Anbajagane}, D. and {Bom}, C.~R. and {Carballo-Bello}, J.~A. and {Carlin}, J.~L. and {Chang}, C. and {Chaturvedi}, A. and {Chiti}, A. and {Choi}, Y. and {Collins}, M.~L.~M. and {Doliva-Dolinsky}, A. and {Ferguson}, P.~S. and {Gruendl}, R.~A. and {James}, D.~J. and {Limberg}, G. and {Navabi}, M. and {Mart{\'\i}nez-Delgado}, D. and {Mart{\'\i}nez-V{\'a}zquez}, C.~E. and {Medina}, G.~E. and {Mutlu-Pakdil}, B. and {Nidever}, D.~L. and {No{\"e}l}, N.~E.~D. and {Riley}, A.~H. and {Sakowska}, J.~D. and {Sand}, D.~J. and {Sharp}, J. and {Stringfellow}, G.~S. and {Tolley}, C. and {Tucker}, D.~L. and {Vivas}, A.~K. and {Delve Collaboration}},
        title = "{A Pride of Satellites in the Constellation Leo? Discovery of the Leo VI Milky Way Satellite Ultra-faint Dwarf Galaxy with DELVE Early Data Release 3}",
      journal = {\apj},
         year = 2025,
        month = feb,
       volume = {979},
       number = {2},
          eid = {176},
        pages = {176},
          doi = {10.3847/1538-4357/ad9b0c},
archivePrefix = {arXiv},
       eprint = {2408.00865},
 primaryClass = {astro-ph.GA},
       adsurl = {https://ui.adsabs.harvard.edu/abs/2025ApJ...979..176T}
}

@ARTICLE{Manwadkar2022,
       author = {{Manwadkar}, Viraj and {Kravtsov}, Andrey V.},
        title = "{Forward-modelling the luminosity, distance, and size distributions of the Milky Way satellites}",
      journal = {\mnras},
         year = 2022,
        month = nov,
       volume = {516},
       number = {3},
        pages = {3944-3971},
          doi = {10.1093/mnras/stac2452},
archivePrefix = {arXiv},
       eprint = {2112.04511},
 primaryClass = {astro-ph.GA},
       adsurl = {https://ui.adsabs.harvard.edu/abs/2022MNRAS.516.3944M}
}

@ARTICLE{Kaplinghat2019,
       author = {{Kaplinghat}, Manoj and {Valli}, Mauro and {Yu}, Hai-Bo},
        title = "{Too big to fail in light of Gaia}",
      journal = {\mnras},
         year = 2019,
        month = nov,
       volume = {490},
       number = {1},
        pages = {231-242},
          doi = {10.1093/mnras/stz2511},
archivePrefix = {arXiv},
       eprint = {1904.04939},
 primaryClass = {astro-ph.GA},
       adsurl = {https://ui.adsabs.harvard.edu/abs/2019MNRAS.490..231K}
}

@ARTICLE{mcdaniel2024,
       author = {{McDaniel}, Alex and {Ajello}, Marco and {Karwin}, Christopher M. and {Di Mauro}, Mattia and {Drlica-Wagner}, Alex and {S{\'a}nchez-Conde}, Miguel A.},
        title = "{Legacy analysis of dark matter annihilation from the Milky Way dwarf spheroidal galaxies with 14 years of Fermi -LAT data}",
      journal = {\prd},
         year = 2024,
        month = mar,
       volume = {109},
       number = {6},
          eid = {063024},
        pages = {063024},
          doi = {10.1103/PhysRevD.109.063024},
archivePrefix = {arXiv},
       eprint = {2311.04982},
 primaryClass = {astro-ph.HE},
       adsurl = {https://ui.adsabs.harvard.edu/abs/2024PhRvD.109f3024M}
}

@ARTICLE{pace2019,
       author = {{Pace}, Andrew B. and {Strigari}, Louis E.},
        title = "{Scaling relations for dark matter annihilation and decay profiles in dwarf spheroidal galaxies}",
      journal = {\mnras},
         year = 2019,
        month = jan,
       volume = {482},
       number = {3},
        pages = {3480-3496},
          doi = {10.1093/mnras/sty2839},
archivePrefix = {arXiv},
       eprint = {1802.06811},
 primaryClass = {astro-ph.GA},
       adsurl = {https://ui.adsabs.harvard.edu/abs/2019MNRAS.482.3480P}
}

@ARTICLE{geha2009,
       author = {{Geha}, Marla and {Willman}, Beth and {Simon}, Joshua D. and {Strigari}, Louis E. and {Kirby}, Evan N. and {Law}, David R. and {Strader}, Jay},
        title = "{The Least-Luminous Galaxy: Spectroscopy of the Milky Way Satellite Segue 1}",
      journal = {\apj},
         year = 2009,
        month = feb,
       volume = {692},
       number = {2},
        pages = {1464-1475},
          doi = {10.1088/0004-637X/692/2/1464},
archivePrefix = {arXiv},
       eprint = {0809.2781},
 primaryClass = {astro-ph},
       adsurl = {https://ui.adsabs.harvard.edu/abs/2009ApJ...692.1464G}
}

@ARTICLE{strigari2008,
       author = {{Strigari}, Louis E. and {Bullock}, James S. and {Kaplinghat}, Manoj and {Simon}, Joshua D. and {Geha}, Marla and {Willman}, Beth and {Walker}, Matthew G.},
        title = "{A common mass scale for satellite galaxies of the Milky Way}",
      journal = {\nat},
         year = 2008,
        month = aug,
       volume = {454},
       number = {7208},
        pages = {1096-1097},
          doi = {10.1038/nature07222},
archivePrefix = {arXiv},
       eprint = {0808.3772},
 primaryClass = {astro-ph},
       adsurl = {https://ui.adsabs.harvard.edu/abs/2008Natur.454.1096S}
}

@ARTICLE{willman2012,
       author = {{Willman}, B. and {Strader}, J.},
        title = "{``Galaxy,'' Defined}",
      journal = {\aj},
         year = 2012,
        month = sep,
       volume = {144},
       number = {3},
          eid = {76},
        pages = {76},
          doi = {10.1088/0004-6256/144/3/76},
archivePrefix = {arXiv},
       eprint = {1203.2608},
 primaryClass = {astro-ph.CO},
       adsurl = {https://ui.adsabs.harvard.edu/abs/2012AJ....144...76W}
}


\appendix


\section{Comparison to Literature Measurements}
\label{sec:appendix}

We compare our uniformly-determined DEIMOS-only measurements presented in this paper to the literature compilation v1.0.6 from \citet{pace2024}.   For this comparison, we focus only on systems identified as satellite galaxies.   In the left panels of Figures~\ref{fig_appendix_mass}, \ref{fig_appendix_feh}, and \ref{fig_appendix_feh_sigma}, we plot our DEIMOS-determined measurements as red symbols, and a linear fit to these data as a red line.   

In the right panels of Figures~\ref{fig_appendix_mass}--\ref{fig_appendix_feh_sigma}, blue symbols are literature values for the same satellite systems included in the DEIMOS sample.   In some cases, these are based on the same DEIMOS dataset, but different data reduction pipelines.   The DEIMOS sample itself is complete for currently known satellite galaxies with Declination $>-40^{\circ}$, with the exception of Antlia\,II and Crater\,II.  There are 12 known satellite galaxies which are not in the DEIMOS archives, shown as green symbols in Figures~\ref{fig_appendix_mass}--\ref{fig_appendix_feh_sigma}.    In these figures, the blue line is a fit to the blue sample, while the green line is a fit to the combined blue and green literature samples.   For direct comparison, we repeat the red line determined from DEIMOS-only measurements in the both panels.

In the left panels of Figures~\ref{fig_appendix_mass}-- \ref{fig_appendix_feh_sigma}, we also compare to selected literature predictions.  This is not intended as a comprehensive comparison, but instead plot recent published predictions which can be directly compared.     From the semi-analytic model GRUMPY, we plot the predicted enclosed dynamical mass from \citet{Kravtsov2023} in Figure~\ref{fig_appendix_mass} and the predicted stellar mass-metallicity relation from \citet{Manwadkar2022} in Figure~\ref{fig_appendix_feh}.   From the semi-analytic model Galacticus \citep{ahvazi2025}, we plot the predicted velocity dispersion versus stellar mass in Figure~\ref{fig_dwarfs}, and the stellar mass-metallicity relation in Figure~\ref{fig_appendix_feh}.

\setcounter{equation}{0}
\setcounter{figure}{0} 
\setcounter{table}{0}
\renewcommand{\thefigure}{A\arabic{figure}}
\renewcommand{\thetable}{A\arabic{table}}

\clearpage
\begin{figure*}
 \includegraphics[width=1.0\textwidth]{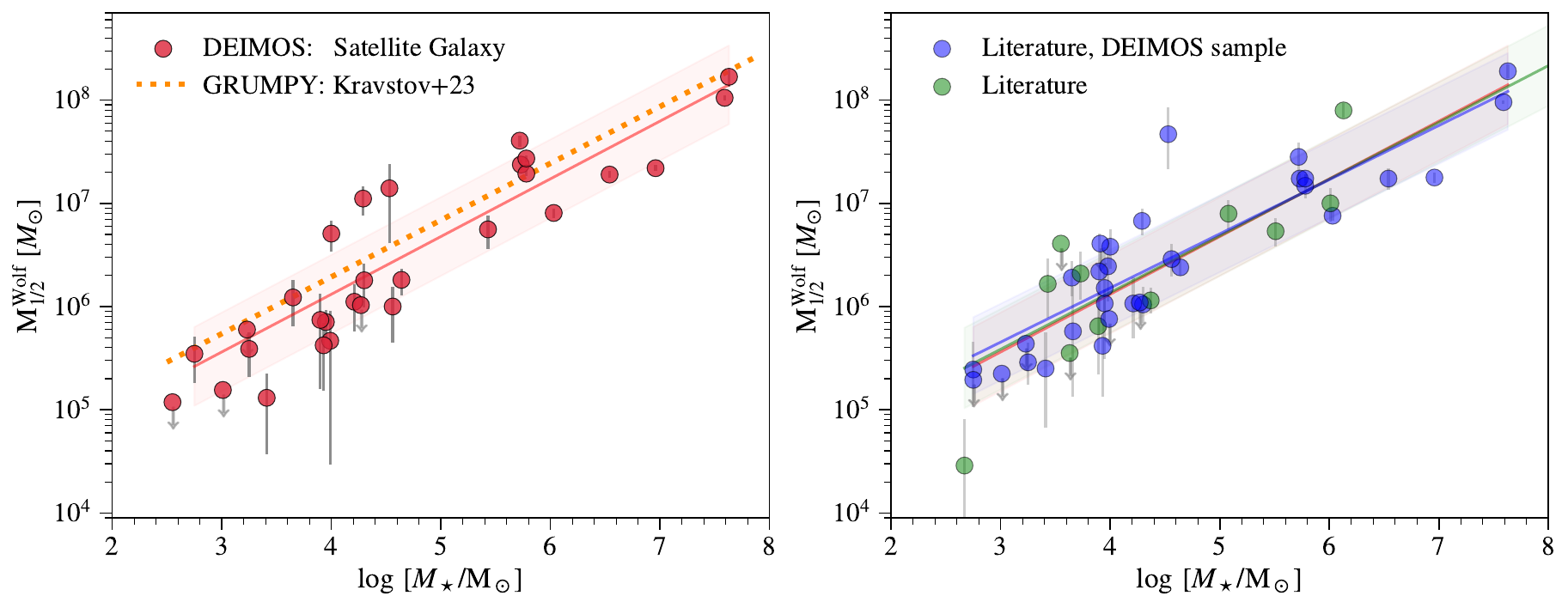}
\caption{Focusing only on satellite galaxies, we plot the enclosed dynamical mass versus stellar mass, $M_{1/2}$ vs.$\log M_{\star}$.   {\it Left:\/}  We include the uniform measurements from the DEIMOS-only sample presented in this paper (red symbols) and fit a linear function (red line).   We also include a comparison to the GRUMPY semi-analytic model \citep[orange dotted line][]{Kravtsov2023}.  {\it Right:\/} Literature values taken from v1.0.6 of \citet{pace2024}. Blue symbols are literature values for the same set of systems presented in the left panel, green are literature values for Milky Way satellites not covered by Keck/DEIMOS.   The red line DEIMOS-only fit is repeated in both panels.  \label{fig_appendix_mass}}
\end{figure*}

\begin{figure*}
 \includegraphics[width=1.0\textwidth]{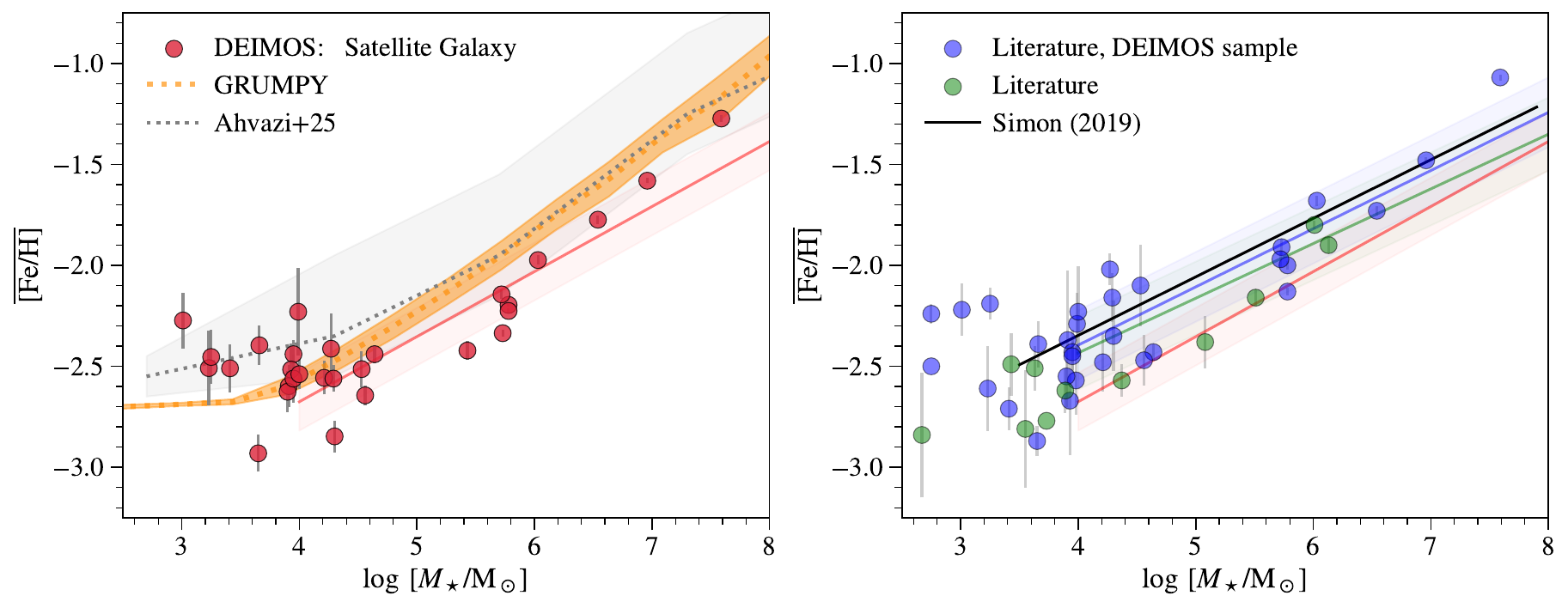}
\caption{Focusing only on satellite galaxies, we plot the stellar metallicity versus stellar mass, $\overline{\rm [Fe/H]}$ vs.~M$_\star$.   Red/blue/green symbols are the same as Figure~\ref{fig_appendix_mass}.   {\it Left:\/}  We compare to the semi-analytic predictions from GRUMPY \citep[orange dotted][]{Manwadkar2022} and Galacticus \citep[gray dotted][]{ahvazi2025}.  Both models suggest a similar flattening below $\log M_{\star}/M_\odot \approx 4$ as observed.  {\it Right:\/}  In comparison to the \citet{pace2024} compilation, the DEIMOS-only stellar mass-metallicity relationship is slightly steeper and the normalization is 0.15\,dex lower. \label{fig_appendix_feh}}
\end{figure*}

\begin{figure*}
 \includegraphics[width=1.0\textwidth]{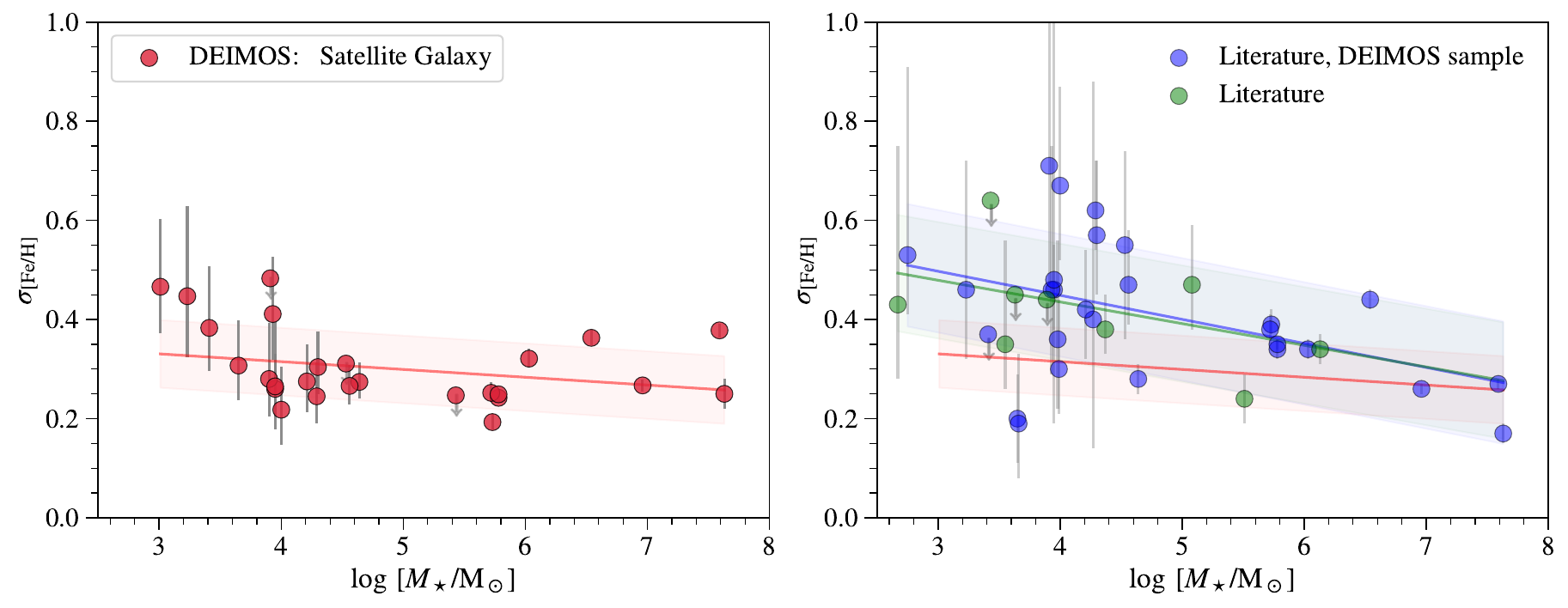}
\caption{Focusing only on satellite galaxies, we plot the stellar metallicity dispersion versus stellar mass, $\sigma_{\rm [Fe/H]}$ vs.~M$_{\star}$.  Red/blue/green symbols are the same as Figure~\ref{fig_appendix_mass}.  {\it Right:\/}  In comparison to the \citet{pace2024} compilation, the DEIMOS-only stellar metallicity spreads are smaller and show less dependence on stellar mass. \label{fig_appendix_feh_sigma}}
\end{figure*}

\clearpage

\begin{longrotatetable}
\begin{center}
\begin{deluxetable*}{rrrrrrrccrrrrrrrrrrrr}
\tabletypesize{\footnotesize}
\tablecaption{Derived Properties of Milky Way Satellites Observed with Keck/DEIMOS\label{table_objects}}
\tablehead{\colhead{Full} & 
		\colhead{System} &
        \colhead{RA} & 
        \colhead{Dec} & 
        \colhead{Dist} & 
	    \colhead{$M_{V}$} & 
	    \colhead{$r_{\rm 1/2}$} & 
        \colhead{Type} &
	    \colhead{$N_*$}&
        \colhead{$v_{\rm sys}$}&
	    \colhead{$\epsilon_{v_{\rm sys}}$} & 
        \colhead{$\sigma_{\rm sys}^{2r_{\rm eff}}$}&
	    \colhead{$\epsilon_{\sigma_{\rm sys}}^{ll}$} & 
	    \colhead{$\epsilon_{\sigma_{\rm sys}}^{ul}$} &
        \colhead{$\epsilon_{\sigma_{\rm sys}}^{95}$}  &
        \colhead{$\overline{\rm [Fe/H]}$}&
	    \colhead{$\epsilon_{\overline{\rm [Fe/H]}}$} & 
        \colhead{$\sigma_{\rm [Fe/H]}$}&
	    \colhead{$\epsilon_{\sigma_{\rm [Fe/H]}}^{ll}$} & 
	    \colhead{$\epsilon_{\sigma_{\rm [Fe/H]}}^{ul}$} &
        \colhead{$\epsilon_{\sigma_{\rm [Fe/H]}}^{95}$} \\        
        \colhead{Name} & 
        \colhead{Name} & 
        \colhead{[deg]} & 
        \colhead{[deg]} &
		\colhead{[kpc]} & 
        \colhead{} & 
        \colhead{[$'$]} & 
        \colhead{} & 
        \colhead{} &
        \colhead{[km/s]} &
        \colhead{[km/s]} &
        \colhead{[km/s]} &
        \colhead{[km/s]} &
        \colhead{[km/s]} &
        \colhead{[km/s]} &
        \colhead{[dex]} &
        \colhead{[dex]}  &
        \colhead{[dex]} &
        \colhead{[dex]}  &        
        \colhead{[dex]} &
        \colhead{[dex]}  \\
        \colhead{(1)} & \colhead{(2)} & \colhead{(3)} & 
        \colhead{(4)} & \colhead{(5)} & \colhead{(6)} &
        \colhead{(7)} &\colhead{(8)} &\colhead{(9)}&
        \colhead{(10)} & \colhead{(11)} & \colhead{(12)} & 
        \colhead{(13)} & \colhead{(14)} & \colhead{(15)}& \colhead{(16)} & \colhead{(17)} &
         \colhead{(18)} & \colhead{(19)} & \colhead{(20)}& \colhead{(21)} \\   
        }
\startdata
 Aquarius III & Aqr3    & 357.2200 & -3.4900  & 85.5  & -2.5  &   0.052 & G      &      11 & -13.2  & 1.0      & -999    &    -999    &   -999    &      3.65 & -2.51 & 0.18    & 0.45       &       0.12 &      0.18 &   -999    \\
 Bootes I     & Boo1    & 210.0200 & 14.5135  & 66.4  & -6.0  &   0.192 & G      &      90 & 101.9  & 0.5      & 3.19   &       0.41 &      0.48 &   -999    & -2.44 & 0.05    & 0.27       &       0.03 &      0.04 &   -999    \\
 Bootes II    & Boo2    & 209.5141 & 12.8553  & 41.7  & -2.9  &   0.038 & G      &      17 & -128.4 & 0.7      & 1.92   &       0.6  &      0.77 &   -999    & -2.51 & 0.12    & 0.38       &       0.09 &      0.12 &   -999    \\
 Bootes III   & Boo3    & 209.3000 & 26.8000  & 46.6  & -5.7  &   0.548 & G      &      16 & 189.9  & 1.8      & 5.27   &       1.67 &      2.1  &   -999    & -2.51 & 0.09    & -999        &    -999    &   -999    &      0.31 \\
 ComaBer      & CB      & 186.7454 & 23.9069  & 42.3  & -4.3  &   0.069 & G      &      82 & 93.5   & 0.6      & 3.33   &       0.46 &      0.58 &   -999    & -2.44 & 0.07    & 0.26       &       0.06 &      0.07 &   -999    \\
 CVn I        & CVn1    & 202.0091 & 33.5521  & 210.9 & -8.7  &   0.436 & G      &     254 & 30.5   & 0.6      & 7.66   &       0.41 &      0.44 &   -999    & -2.34 & 0.02    & 0.19       &       0.02 &      0.02 &   -999    \\
 CVn II       & CVn2    & 194.2927 & 34.3226  & 160.0 & -5.2  &   0.07  & G      &      25 & -131.8 & 1.2      & 5.28   &       0.96 &      1.25 &   -999    & -2.85 & 0.08    & 0.30       &       0.05 &      0.07 &   -999    \\
 Draco        & Dra     & 260.0685 & 57.9185  & 81.5  & -8.9  &   0.229 & G      &     995 & -292.2 & 0.3      & 9.55   &       0.25 &      0.26 &   -999    & -2.20 & 0.01    & 0.24       &       0.01 &      0.01 &   -999    \\
 Draco II     & Dra2    & 238.2000 & 64.5700  & 21.6  & -0.8  &   0.019 & G      &      25 & -342.3 & 0.7      & -999    &    -999    &   -999    &      2.64 & -999   & -999     & -999        &    -999    &   -999    &   -999    \\
 Eridanus     & Eri     & 66.1853  & -21.1876 & 84.7  & -5.4  &   0.016 & GC     &      24 & -18.4  & 0.3      & -999    &    -999    &   -999    &      0.89 & -1.29 & 0.05    & -999        &    -999    &   -999    &      0.23 \\
 Eridanus 4   & Eri4    & 76.4380  & -9.5150  & 69.8  & -3.5  &   0.065 & G      &      19 & -34.4  & 1.3      & 4.55   &       0.89 &      1.15 &   -999    & -2.93 & 0.09    & 0.31       &       0.07 &      0.09 &   -999    \\
 Fornax       & For     & 39.9583  & -34.4997 & 142.6 & -13.4 &   0.825 & G      &     663 & 53.9   & 0.5      & 11.71  &       0.33 &      0.36 &   -999    & -1.27 & 0.02    & 0.38       &       0.01 &      0.01 &   -999    \\
 Herc         & Herc    & 247.7722 & 12.7852  & 130.6 & -5.8  &   0.213 & G      &      43 & 45.3   & 0.6      & 2.25   &       0.61 &      0.67 &   -999    & -2.64 & 0.05    & 0.27       &       0.04 &      0.04 &   -999    \\
 Hydra 2      & Hyd2    & 185.4251 & -31.9860 & 151.4 & -5.1  &   0.066 & G      &      11 & 304.4  & 1.1      & -999    &    -999    &   -999    &      4.11 & -2.41 & 0.17    & -999        &    -999    &   -999    &   -999    \\
 Koposov 2    & K2      & 119.5715 & 26.2574  & 24.0  & -0.9  &   0.004 & U      &      13 & 108.2  & 1.3      & -999    &    -999    &   -999    &      5.23 & -999   & -999     & -999        &    -999    &   -999    &   -999    \\
 Leo I        & Leo1    & 152.1146 & 12.3059  & 258.2 & -11.8 &   0.274 & G      &     686 & 285.9  & 0.4      & 9.29   &       0.26 &      0.26 &   -999    & -1.58 & 0.01    & 0.27       &       0.01 &      0.01 &   -999    \\
 Leo II       & Leo2    & 168.3627 & 22.1529  & 208.9 & -9.5  &   0.153 & G      &     253 & 78.8   & 0.5      & 7.53   &       0.36 &      0.39 &   -999    & -1.97 & 0.02    & 0.32       &       0.02 &      0.02 &   -999    \\
 Leo IV       & Leo4    & 173.2405 & -0.5431  & 151.4 & -5.0  &   0.111 & G      &      24 & 129.8  & 0.9      & 3.29   &       0.66 &      0.89 &   -999    & -2.56 & 0.08    & 0.28       &       0.06 &      0.07 &   -999    \\
 Leo V        & Leo5    & 172.7857 & 2.2194   & 169.0 & -4.4  &   0.049 & G      &      10 & 170.8  & 1.5      & 3.37   &       1.04 &      1.72 &   -999    & -2.23 & 0.22    & -999        &    -999    &   -999    &   -999    \\
 Leo T        & LeoT    & 143.7292 & 17.0482  & 413.1 & -8.0  &   0.167 & G      &      42 & 37.1   & 1.1      & 6.05   &       0.86 &      1.07 &   -999    & -2.42 & 0.05    & -999        &    -999    &   -999    &      0.25 \\
 NGC 1904     & N1904   & 81.0496  & -24.5247 & 13.1  & -7.7  &   0.003 & GC     &      61 & 206.0  & 0.7      & 5.22   &       0.51 &      0.57 &   -999    & -1.59 & 0.01    & -999        &    -999    &   -999    &      0.06 \\
 NGC 2419     & N2419   & 114.5354 & 38.8819  & 88.5  & -9.4  &   0.022 & GC     &      86 & -20.9  & 0.5      & 4.69   &       0.38 &      0.43 &   -999    & -2.02 & 0.02    & -999      &       -999 &      -999 &   -999    \\
 NGC 0288     & N288    & 13.1977  & -26.5899 & 9.0   & -6.8  &   0.006 & GC     &     473 & -43.2  & 0.2      & 3.47   &       0.16 &      0.17 &   -999    & -1.23 & 0.01    & -999        &    -999    &   -999    &      0.03 \\
 NGC 4590     & N4590   & 189.8667 & -26.7430 & 10.4  & -7.2  &   0.004 & GC     &     103 & -93.6  & 0.4      & 3.31   &       0.25 &      0.28 &   -999    & -2.33 & 0.01    & -999        &    -999    &   -999    &      0.03 \\
 NGC 5024     & N5024   & 198.2301 & 18.1691  & 18.5  & -8.7  &   0.006 & GC     &      33 & -64.8  & 0.8      & 3.77   &       0.57 &      0.66 &   -999    & -1.95 & 0.02    & -999        &    -999    &   -999    &      0.12 \\
 NGC 5053     & N5053   & 199.1124 & 17.6977  & 17.5  & -6.3  &   0.012 & GC     &      83 & 42.5   & 0.2      & -999    &    -999    &   -999    &      1.2  & -2.34 & 0.02    & -999        &    -999    &   -999    &      0.07 \\
 NGC 5272     & N5272   & 205.5468 & 28.3755  & 10.2  & -8.7  &   0.003 & GC     &      44 & -147.0 & 0.7      & 3.85   &       0.46 &      0.53 &   -999    & -1.54 & 0.02    & 0.08       &       0.02 &      0.02 &   -999    \\
 NGC 5634     & N5634   & 217.4053 & -5.9764  & 25.9  & -7.7  &   0.005 & GC     &      42 & -18.3  & 0.6      & 3.40   &       0.4  &      0.47 &   -999    & -1.80 & 0.01    & -999        &    -999    &   -999    &      0.05 \\
 NGC 5897     & N5897   & 229.3517 & -21.0101 & 12.6  & -7.3  &   0.008 & GC     &     152 & 101.4  & 0.3      & 2.93   &       0.21 &      0.25 &   -999    & -2.01 & 0.01    & -999        &    -999    &   -999    &      0.04 \\
 NGC 5904     & N5904   & 229.6406 & 2.0827   & 7.5   & -8.5  &   0.004 & GC     &      50 & 55.2   & 0.7      & 4.49   &       0.47 &      0.56 &   -999    & -1.24 & 0.02    & -999        &    -999    &   -999    &      0.04 \\
 NGC 6205     & N6205   & 250.4235 & 36.4613  & 7.4   & -8.6  &   0.003 & GC     &      72 & -245.8 & 0.7      & 5.74   &       0.49 &      0.6  &   -999    & -1.50 & 0.01    & -999        &    -999    &   -999    &      0.04 \\
 NGC 6218     & N6218   & 251.8105 & -1.9478  & 5.1   & -7.0  &   0.003 & GC     &      11 & -40.6  & 0.8      & -999    &    -999    &   -999    &      3.3  & -999   & -999     & -999        &    -999    &   -999    &   -999    \\
 NGC 6229     & N6229   & 251.7452 & 47.5278  & 30.1  & -8.1  &   0.004 & GC     &      31 & -139.1 & 0.6      & 2.38   &       0.45 &      0.59 &   -999    & -1.20 & 0.04    & -999        &    -999    &   -999    &      0.14 \\
 NGC 6254     & N6254   & 254.2875 & -4.0993  & 5.1   & -7.8  &   0.003 & GC     &      12 & 73.5   & 1.9      & 6.18   &       1.27 &      1.85 &   -999    & -1.46 & 0.05    & -999        &    -999    &   -999    &   -999    \\
 NGC 6341     & N6341   & 259.2803 & 43.1365  & 8.5   & -8.2  &   0.002 & GC     &     110 & -120.0 & 0.5      & 4.61   &       0.36 &      0.4  &   -999    & -2.35 & 0.01    & -999        &    -999    &   -999    &      0.09 \\
 NGC 6366     & N6366   & 261.9347 & -5.0766  & 3.4   & -6.0  &   0.004 & GC     &      21 & -120.6 & 0.8      & 3.28   &       0.57 &      0.76 &   -999    & -0.63 & 0.05    & -999        &    -999    &   -999    &      0.14 \\
 NGC 6624     & N6624   & 275.9190 & -30.3613 & 8.0   & -7.4  &   0.002 & GC     &      69 & 55.4   & 0.4      & 3.29   &       0.31 &      0.36 &   -999    & -0.67 & 0.02    & -999        &    -999    &   -999    &      0.04 \\
 NGC 6656     & N6656   & 279.1008 & -23.9034 & 3.3   & -8.6  &   0.003 & GC     &      74 & -146.3 & 0.9      & 7.05   &       0.62 &      0.69 &   -999    & -1.61 & 0.02    & -999        &    -999    &   -999    &      0.14 \\
 NGC 6715     & N6715   & 283.7639 & -30.4799 & 26.3  & -10.0 &   0.004 & GC     &      15 & 142.2  & 1.6      & 6.04   &       1.08 &      1.45 &   -999    & -1.33 & 0.03    & -999        &    -999    &   -999    &      0.13 \\
 NGC 6779     & N6779   & 289.1479 & 30.1845  & 10.4  & -7.7  &   0.003 & GC     &      74 & -133.6 & 0.6      & 4.70   &       0.41 &      0.48 &   -999    & -1.93 & 0.02    & -999        &    -999    &   -999    &      0.06 \\
 NGC 6838     & N6838   & 298.4421 & 18.7784  & 4.0   & -6.6  &   0.003 & GC     &      17 & -21.8  & 0.6      & 2.07   &       0.47 &      0.58 &   -999    & -0.73 & 0.05    & -999        &    -999    &   -999    &      0.13 \\
 NGC 6864     & N6864   & 301.5202 & -21.9222 & 20.5  & -8.5  &   0.003 & GC     &      36 & -186.3 & 0.7      & 3.66   &       0.48 &      0.56 &   -999    & -1.08 & 0.03    & -999        &    -999    &   -999    &      0.14 \\
 NGC 7006     & N7006   & 315.3722 & 16.1871  & 39.3  & -7.5  &   0.004 & GC     &      22 & -383.1 & 0.8      & 3.25   &       0.55 &      0.69 &   -999    & -1.47 & 0.02    & -999        &    -999    &   -999    &      0.07 \\
 NGC 7078     & N7078   & 322.4932 & 12.1668  & 10.7  & -9.2  &   0.002 & GC     &     246 & -105.5 & 0.4      & 6.18   &       0.3  &      0.33 &   -999    & -2.33 & 0.01    & -999        &    -999    &   -999    &      0.05 \\
 NGC 7089     & N7089   & 323.3626 & -0.8233  & 11.7  & -9.1  &   0.003 & GC     &     200 & -2.6   & 0.5      & 6.81   &       0.34 &      0.41 &   -999    & -1.53 & 0.01    & -999        &    -999    &   -999    &      0.04 \\
 NGC 7099     & N7099   & 325.0918 & -23.1791 & 8.5   & -7.4  &   0.003 & GC     &     146 & -185.1 & 0.4      & 4.01   &       0.28 &      0.3  &   -999    & -2.27 & 0.01    & -999        &    -999    &   -999    &      0.05 \\
 NGC 7492     & N7492   & 347.1102 & -15.6108 & 24.4  & -5.9  &   0.009 & GC     &      21 & -175.4 & 0.5      & 1.75   &       0.46 &      0.51 &   -999    & -1.86 & 0.02    & -999        &    -999    &   -999    &      0.08 \\
 Palomar 13   & Pal13   & 346.6858 & 12.7712  & 23.5  & -3.1  &   0.008 & GC     &      57 & 26.0   & 0.3      & -999    &    -999    &   -999    &      1.52 & -1.64 & 0.03    & -999        &    -999    &   -999    &      0.09 \\
 Palomar 14   & Pal14   & 242.7544 & 14.9584  & 73.6  & -5.3  &   0.029 & GC     &      25 & 73.1   & 0.4      & -999    &    -999    &   -999    &      1.56 & -1.50 & 0.06    & 0.21       &       0.06 &      0.06 &   -999    \\
 Palomar 2    & Pal2    & 71.5248  & 31.3817  & 26.2  & -8.3  &   0.009 & GC     &      15 & -111.7 & 1.3      & 4.45   &       0.9  &      1.29 &   -999    & -999   & -999     & -999        &    -999    &   -999    &   -999    \\
 Palomar 5    & Pal5    & 229.0220 & -0.1113  & 21.9  & -4.9  &   0.02  & GC     &      77 & -57.7  & 0.6      & 3.21   &       0.69 &      0.71 &   -999    & -999   & -999     & -999        &    -999    &   -999    &   -999    \\
 Palomar 7    & Pal7    & 272.6844 & -7.2076  & 4.5   & -6.7  &   0.003 & GC     &      15 & 157.4  & 1.7      & 5.66   &       1.18 &      1.72 &   -999    & -999   & -999     & -999        &    -999    &   -999    &   -999    \\
 Pegasus 3    & Peg3    & 336.1000 & 5.4100   & 214.8 & -4.2  &   0.104 & G      &      15 & -260.0 & 1.1      & 2.76   &       0.91 &      1.24 &   -999    & -2.63 & 0.10    & 0.28       &       0.08 &      0.11 &   -999    \\
 Pegasus 4    & Peg4    & 328.5390 & 26.6200  & 90.0  & -4.2  &   0.042 & G      &      21 & -271.4 & 1.1      & 3.33   &       0.86 &      1.19 &   -999    & -2.52 & 0.13    & 0.41       &       0.09 &      0.12 &   -999    \\
 Sculptor     & Scl     & 15.0183  & -33.7186 & 84.0  & -10.8 &   0.273 & G      &     379 & 110.7  & 0.5      & 8.66   &       0.33 &      0.33 &   -999    & -1.77 & 0.02    & 0.36       &       0.02 &      0.02 &   -999    \\
 Segue 1      & Seg1    & 151.7504 & 16.0756  & 22.9  & -1.3  &   0.024 & G      &      53 & 203.1  & 0.9      & 3.97   &       0.86 &      0.97 &   -999    & -999   & -999     & -999        &    -999    &   -999    &   -999    \\
 Segue 2      & Seg2    & 34.8226  & 20.1625  & 36.5  & -1.9  &   0.04  & G      &      29 & -41.2  & 0.4      & -999    &    -999    &   -999    &      2.06 & -2.27 & 0.14    & 0.47       &       0.09 &      0.14 &   -999    \\
 Segue 3      & Seg3    & 320.3795 & 19.1178  & 16.9  & -0.1  &   0.002 & GC     &      17 & -165.5 & 0.8      & -999    &    -999    &   -999    &      3.27 & -999   & -999     & -999        &    -999    &   -999    &   -999    \\
 Sextans      & Sext    & 153.2628 & -1.6133  & 85.9  & -8.7  &   0.569 & G      &     239 & 221.9  & 0.6      & 8.77   &       0.47 &      0.49 &   -999    & -2.14 & 0.02    & 0.25       &       0.02 &      0.02 &   -999    \\
 Sgr          & Sgr     & 284.0952 & -30.5499 & 26.3  & -13.5 &   2.608 & G      &      60 & 142.7  & 1.1      & 8.31   &       0.74 &      0.86 &   -999    & -0.35 & 0.03    & 0.16       &       0.03 &      0.03 &   -999    \\
 Sgr 2        & Sgr2    & 298.1700 & -22.0700 & 64.0  & -5.8  &   0.034 & GC     &      28 & -176.4 & 0.6      & 2.32   &       0.43 &      0.53 &   -999    & -2.31 & 0.04    & -999        &    -999    &   -999    &      0.17 \\
 Terzan 5     & Ter5    & 267.0021 & -24.7792 & 6.6   & -9.1  &   0.002 & GC     &      15 & -85.4  & 2.7      & 10.27  &       1.74 &      2.45 &   -999    & -999   & -999     & -999        &    -999    &   -999    &   -999    \\
 UMa 1        & UMa1    & 158.7706 & 51.9480  & 97.3  & -5.1  &   0.234 & G      &      36 & -58.5  & 1.4      & 7.16   &       1.04 &      1.21 &   -999    & -2.56 & 0.07    & 0.24       &       0.05 &      0.07 &   -999    \\
 UMa 2        & UMa2    & 132.8726 & 63.1335  & 34.7  & -4.4  &   0.139 & G      &      64 & -118.0 & 1.1      & 6.30   &       0.95 &      1.06 &   -999    & -2.54 & 0.07    & 0.22       &       0.07 &      0.09 &   -999    \\
 Ursa Minor   & UMi     & 227.2420 & 67.2221  & 70.2  & -8.9  &   0.373 & G      &     827 & -246.1 & 0.3      & 8.86   &       0.24 &      0.27 &   -999    & -2.23 & 0.01    & 0.25       &       0.01 &      0.01 &   -999    \\
 Unions 1     & UNI1    & 174.7080 & 31.0711  & 10.0  & 2.2   &   0.003 & U      &      10 & 89.6   & 0.9      & -999    &    -999    &   -999    &      4.35 & -999   & -999     & -999        &    -999    &   -999    &   -999    \\
 Willman 1    & W1      & 162.3436 & 51.0501  & 38.0  & -2.5  &   0.027 & G      &      47 & -12.5  & 0.9      & 3.97   &       0.72 &      0.88 &   -999    & -2.45 & 0.14    & -999        &    -999    &   -999    &   -999    \\
\enddata
\tablecomments{Literature and computed properties of Milky Way stellar satellites included in this work.  (1)-(2) Name and abbreviated name of system, (3)-(7) position, distance, absolute magnitude and half-light radius taken from \citet{pace2024}. (9)  number of member stars identified.    (10)-(11) The systematic velocity and associated error determined for the system.  (12)-(15)  The velocity dispersion and associated error determined for the system.  (16)-(17)  The mean [Fe/H] and error determined for the system. (18)-(21)  The internal [Fe/H] spread and error determined for the system.  The full contents of this table (67 rows) is available in a machine-readable format. }
\end{deluxetable*}
\end{center}
\end{longrotatetable}

\end{document}